%
%
%
%
%
%
%
%
%
%
%
%
%
\font\twelverm = cmr10  at 12truept
\font\ninerm   = cmr9
\font\twelvei  = cmmi10 at 12truept
\font\ninei    = cmmi9
\font\twelvesy = cmsy10 at 12truept
\font\ninesy   = cmsy9
\font\twelvebf = cmbx10 at 12truept
\font\ninebf   = cmbx9
\font\twelvex  = cmex10 at 12truept
\font\twelveit = cmti10 at 12truept
\font\twelvesl = cmsl10 at 12truept
\font\twelvett = cmtt10 at 12truept
%
%
  at 14.4truept    
\font\caps  = cmcsc10 at 12.0truept    
%
%
\def\twelvepoint{   \def\rm{ \fam0 \twelverm }
  \textfont0=\twelverm \scriptfont0=\ninerm  \scriptscriptfont0=\sevenrm
  \textfont1=\twelvei  \scriptfont1=\ninei   \scriptscriptfont1=\seveni
  \textfont2=\twelvesy \scriptfont2=\ninesy  \scriptscriptfont2=\sevensy
  \textfont3=\twelvex  \scriptfont3=\twelvex \scriptscriptfont3=\twelvex
  \textfont\itfam=\twelveit         \def\it{ \fam\itfam \twelveit }
  \textfont\slfam=\twelvesl         \def\sl{ \fam\slfam \twelvesl }
  \textfont\ttfam=\twelvett         \def\tt{ \fam\ttfam \twelvett }
  \textfont\bffam=\twelvebf  \scriptfont\bffam=\ninebf
  \scriptscriptfont\bffam=\sevenbf  \def\bf{ \fam\bffam \twelvebf }
  \normalbaselineskip=16truept
  \normalbaselines \rm   }
%
\twelvepoint
\def\makefootline{\baselineskip=36pt \line{\the\footline}}
%
%
%
%
\newcount\chapnum
\chapnum = 0
\newcount\secnum
\secnum = 0
\newcount\subsecnum
\subsecnum = 0
\newcount\eqnum
\eqnum = 0
%
%
\hsize    6.9in                      
\vsize    8.9in                      
\hoffset  0.0in
\voffset  0.0in
\parskip  5pt plus 1pt               
\def\chskipt{\vskip 20pt plus 0pt minus 0pt }
\def\chskipl{\vskip 5.5pt plus 0pt minus 0pt}
\def\secskipt{\vskip 6pt plus 2pt minus 1pt }

\def\ctrline{\centerline}
\def\title#1{\ctrline {\bf #1} \medskip }
\def\author#1{  \ctrline{ \caps {#1} }  \smallskip
    \ctrline{ \it {Max-Planck-Institut f\"ur Physik und Astrophysik} }
    \ctrline{ \it {Institut f\"ur Astrophysik}                       }
    \ctrline{ \it {Karl-Schwarzschild-Str. 1}                        }
    \ctrline{ \it {D-8046 Garching b. M\"unchen, FRG}                } }
%

%

%

%

%

%
%

%

%

%

%

%

%

%
%
%
%
%
%
%
%
%
%
%
\def\chpreset{ \global \secnum=0     \global \eqnum=0 }
\def\secreset{ \global \subsecnum=0 }
\def\chap#1{ \global \advance\chapnum by 1  \chpreset \chskipt
             \line{\bf \the\chapnum . #1\hfill}
             \penalty 100000 \chskipl \penalty 100000 }
\def\sec#1{ \global \advance\secnum by 1  \secreset \secskipt
    \line{\bf \the\chapnum .\the\secnum{ }{ }#1\hfill}
             \penalty 100000  }
\def\subsec#1{ \global \advance\subsecnum by 1  \secskipt
    \line{\bf \the\chapnum .\the\secnum .\the\subsecnum{ }{ }#1\hfill}
             \penalty 100000  }
\def\eqname#1{ \global \advance\eqnum by 1
  \xdef#1{(\the\chapnum .\the\eqnum )} (\the\chapnum .\the\eqnum ) }
\def\enum{ \global \advance\eqnum by 1 (\the\chapnum .\the\eqnum ) }
\def\ueber#1#2{{\setbox0=\hbox{$#1$}%
  \setbox1=\hbox to\wd0{\hss$\scriptscriptstyle #2$\hss}%
  \offinterlineskip
  \vbox{\box1\box0}}{}}
\hsize 6.5in
\vsize 9.5in
\hoffset  0.00in
\voffset -0.0in
\vskip 1.0in
\baselineskip=14pt

\font\klein=cmr7

\font\mathklein=cmmi5

\font\kleinind=cmr5

\overfullrule=0pt
{\catcode`\@=11
\gdef\SchlangeUnter#1#2{\lower2pt\vbox{\baselineskip 0pt \lineskip0pt
  \ialign{$\m@th#1\hfil##\hfil$\crcr#2\crcr\sim\crcr}}}
}

\parskip=5pt
\def\ref#1{\lbrack #1\rbrack}
\def\eck#1{\left\lbrack #1 \right\rbrack}
\def\rund#1{\left( #1 \right)}
\def\ave#1{\langle #1 \rangle}

\def\ueber#1#2{{\setbox0=\hbox{$#1$}%
  \setbox1=\hbox to\wd0{\hss$\scriptscriptstyle #2$\hss}%
  \offinterlineskip
\def\underline{\underline}
\def\x{x = x^{(m)}}
\def\ax{\underline{x} = \underline{x}^{(m-1)}}
\vbox{\box1\kern0.4mm\box0}}{}}
\def\ref#1{\lbrack #1\rbrack}
\def\eck#1{\left\lbrack #1 \right\rbrack}
\def\rund#1{\left( #1 \right)}
\def\ave#1{\langle #1 \rangle}

\def\ueber#1#2{{\setbox0=\hbox{$#1$}%
  \setbox1=\hbox to\wd0{\hss$\scriptscriptstyle #2$\hss}%
  \offinterlineskip
  \vbox{\box1\kern0.4mm\box0}}{}}
\parindent=12pt
\overfullrule=0pt
%
%
%
%
\font\twelverm = cmr10 scaled\magstep1 \font\tenrm = cmr10
       \font\sevenrm = cmr7
\font\twelvei = cmmi10 scaled\magstep1
       \font\teni = cmmi10 \font\seveni = cmmi7
\font\twelveit = cmti10 scaled\magstep1 
       
\font\twelvesy = cmsy10 scaled\magstep1
       \font\tensy = cmsy10 \font\sevensy = cmsy7
\font\twelvebf = cmbx10 scaled\magstep1 \font\tenbf = cmbx10
       \font\sevenbf = cmbx7
\font\twelvesl = cmsl10 scaled\magstep1
\font\twelveit = cmti10 scaled\magstep1
\font\twelvett = cmtt10 scaled\magstep1
\font\klein=cmr10

\font\gross=cmbx12 at 1132462sp

 at 1132462sp
\font\mathklein=cmmi9

\font\kleinind=cmr7
\font\grossind=cmbx10 at 12truept

%
\textfont0 = \twelverm \scriptfont0 = \twelverm
       \scriptscriptfont0 = \tenrm
       \def\rm{\fam0 \twelverm}
\textfont1 = \twelvei \scriptfont1 = \twelvei
       \scriptscriptfont1 = \teni
       
\textfont2 = \twelvesy \scriptfont2 = \twelvesy
       \scriptscriptfont2 = \tensy
       
\newfam\itfam \def\it{\fam\itfam \twelveit} \textfont\itfam=\twelveit
\newfam\slfam \def\sl{\fam\slfam \twelvesl} \textfont\slfam=\twelvesl
\newfam\bffam \def\bf{\fam\bffam \twelvebf} \textfont\bffam=\twelvebf
       \scriptfont\bffam=\twelvebf \scriptscriptfont\bffam=\tenbf
\newfam\ttfam \def\tt{\fam\ttfam \twelvett} \textfont\ttfam=\twelvett
\rm
\hoffset=0.00in
\hsize=5.8in
\vsize=9in
\baselineskip=16pt
%
%
\dimen1=\baselineskip \multiply\dimen1 by 3 \divide\dimen1 by 4
\dimen2=\dimen1 \divide\dimen2 by 2
%
\nopagenumbers
%
\def\title#1 {\centerline{\gross \textfont1=\gross \textfont0=\gross
\scriptfont1=\gross \scriptfont0=\grossind #1}  }
%
\def\begfig#1 {\midinsert \vskip #1}
\def\figure#1#2{ \baselineskip=12pt {
 \noindent Figure #1: #2}}
\def\endfig {\endinsert}

\def\table#1#2{ \bigskip \baselineskip=12pt {\klein \textfont1=\mathklein
\textfont0=\mathklein
\scriptfont1=\mathklein \scriptfont0=\kleinind \noindent \centerline{TABLE #1}
\hfill \vskip 8pt \noindent{  #2}
\vskip 8pt}  }
\def\endtab{\endinsert \bigskip}
\def\begtabsmall{\midinsert
\klein \textfont1=\mathklein \textfont0=\klein \scriptfont1=\mathklein
\scriptfont0=\kleinind }
%
\def\heading#1{\bigskip \vbox to \dimen1 {\vfill}
     \centerline{\bf #1}
     \vskip \dimen1}
%
\newcount\sectcount
\newcount\subcount
\newcount\subsubcount
\global\sectcount=0
\global\subcount=0
\global\subsubcount=0
\def\section#1{\bigskip   \vbox to \dimen1 {\vfill}
    \global\advance\sectcount by 1
    \centerline{\bf \the\sectcount.\ \ {#1}}
    \global\subcount=0  
    \global\subsubcount=0  
    \vskip \dimen1}
%
\def\subsection#1{\global\advance\subcount by 1
    \vskip \parskip  \vskip \dimen2
    \centerline{{\it \the\sectcount.\the\subcount.\ \ #1}}
    \global\subsubcount=0  
    \vskip \dimen2}
%
\def\subsubsection#1{\global\advance\subsubcount by 1
    \vskip \parskip  \vskip \dimen2
    \centerline{{\it \the\sectcount.\the\subcount.\the\subsubcount.\ \ #1}}
    \vskip \dimen2}
%
%
\def\refindent{\advance\leftskip by 24pt \parindent=-24pt}
%
\def\journal#1#2#3#4#5{{\refindent
                      {#1}        
                      {#2},       
                      {#3},       
                      {#4},       
                      {#5}        
                      \par }}
%
\def\infuture#1#2#3#4{{\refindent
                  {#1}         
                  {#2},        
                  {#3},        
                  {#4}         
                  \par }}
%
\def\inbook#1#2#3#4#5#6#7{{\refindent
                         {#1}         
                         {#2},        
                      in {#3},  
                     ed. {#4}         
                        ({#5}:        
                         {#6}),       
                       p.{#7}         
                         \par }}
%

%

%

%

%

%

%
\def\etal{{\it et al.\/\ }}
\def\eg{{\it e.g.\/}}

%
%
%

\newcount\notenumber
\notenumber=0
\def\hline{\vskip 3pt \hrule\vskip 5pt}
\def\note{\global\advance\notenumber by 1
	  \footnote{$^{\the\notenumber}$}}
\def\apjnote#1{\global\advance \notenumber by 1
	       $^{\the\notenumber}$ \kern -.6em
               \vadjust{\midinsert
                        \hbox to \hsize{\hrulefill}
                        $^{\the\notenumber}${#1} \hfill \break
                        \hbox to \hsize{\hrulefill}
                        \endinsert
                       }
              }

\def\PsfigVersion{1.10}
\def\setDriver{\DvipsDriver} 
\ifx\undefined\psfig\else\endinput\fi
%

\let\LaTeXAtSign=\@
\let\@=\relax
\edef\psfigRestoreAt{\catcode`\@=\number\catcode`@\relax}
\catcode`\@=11\relax
\newwrite\@unused
\def\ps@typeout#1{{\let\protect\string\immediate\write\@unused{#1}}}

\def\DvipsDriver{
	\ps@typeout{psfig/tex \PsfigVersion -dvips}
\def\PsfigSpecials{\DvipsSpecials} 	\def\ps@dir{/}
\def\ps@predir{} }
\def\OzTeXDriver{
	\ps@typeout{psfig/tex \PsfigVersion -oztex}
	\def\PsfigSpecials{\OzTeXSpecials}
	\def\ps@dir{:}
	\def\ps@predir{:}
	\catcode`\^^J=5
}


\def\figurepath{./:}

\def\DoPaths#1{\expandafter\EachPath#1\stoplist}
\def\leer{}
\def\EachPath#1:#2\stoplist{
  \ExistsFile{#1}{\SearchedFile}
  \ifx#2\leer
  \else
    \expandafter\EachPath#2\stoplist
  \fi}
%
%
\def\ps@dir{/}
\def\ExistsFile#1#2{%
   \openin1=\ps@predir#1\ps@dir#2
   \ifeof1
       \closein1
   \else
       \closein1
        \ifx\ps@founddir\leer
           \edef\ps@founddir{#1}
        \fi
   \fi}
%
%
\def\get@dir#1{%
  \def\ps@founddir{}
  \def\SearchedFile{#1}
  \DoPaths\figurepath
}

%
%
\def\@nnil{\@nil}
\def\@empty{}
\def\@psdonoop#1\@@#2#3{}
\def\@psdo#1:=#2\do#3{\edef\@psdotmp{#2}\ifx\@psdotmp\@empty \else
    \expandafter\@psdoloop#2,\@nil,\@nil\@@#1{#3}\fi}
\def\@psdoloop#1,#2,#3\@@#4#5{\def#4{#1}\ifx #4\@nnil \else
       #5\def#4{#2}\ifx #4\@nnil \else#5\@ipsdoloop #3\@@#4{#5}\fi\fi}
\def\@ipsdoloop#1,#2\@@#3#4{\def#3{#1}\ifx #3\@nnil
       \let\@nextwhile=\@psdonoop \else
      #4\relax\let\@nextwhile=\@ipsdoloop\fi\@nextwhile#2\@@#3{#4}}
\def\@tpsdo#1:=#2\do#3{\xdef\@psdotmp{#2}\ifx\@psdotmp\@empty \else
    \@tpsdoloop#2\@nil\@nil\@@#1{#3}\fi}
\def\@tpsdoloop#1#2\@@#3#4{\def#3{#1}\ifx #3\@nnil
       \let\@nextwhile=\@psdonoop \else
      #4\relax\let\@nextwhile=\@tpsdoloop\fi\@nextwhile#2\@@#3{#4}}
%
\ifx\undefined\fbox
\newdimen\fboxrule
\newdimen\fboxsep
\newdimen\ps@tempdima
\newbox\ps@tempboxa
\fboxsep = 3pt
\fboxrule = .4pt
\long\def\fbox#1{\leavevmode\setbox\ps@tempboxa\hbox{#1}\ps@tempdima\fboxrule
    \advance\ps@tempdima \fboxsep \advance\ps@tempdima \dp\ps@tempboxa
   \hbox{\lower \ps@tempdima\hbox
  {\vbox{\hrule height \fboxrule
          \hbox{\vrule width \fboxrule \hskip\fboxsep
          \vbox{\vskip\fboxsep \box\ps@tempboxa\vskip\fboxsep}\hskip
                 \fboxsep\vrule width \fboxrule}
                 \hrule height \fboxrule}}}}
\fi
%
%
\newread\ps@stream
\newif\ifnot@eof       
\newif\if@noisy        
\newif\if@atend        
\newif\if@psfile       
%
%
{\catcode`\%=12\global\gdef\epsf@start{
\def\epsf@PS{PS}
\def\epsf@getbb#1{%
%
%
\openin\ps@stream=\ps@predir#1
\ifeof\ps@stream\ps@typeout{Error, File #1 not found}\else
%
%
   {\not@eoftrue \chardef\other=12
    \def\do##1{\catcode`##1=\other}\dospecials \catcode`\ =10
    \loop
       \if@psfile
	  \read\ps@stream to \epsf@fileline
       \else{
	  \obeyspaces
          \read\ps@stream to \epsf@tmp\global\let\epsf@fileline\epsf@tmp}
       \fi
       \ifeof\ps@stream\not@eoffalse\else
%
%
       \if@psfile\else
       \expandafter\epsf@test\epsf@fileline:. \\%
       \fi
%
%
          \expandafter\epsf@aux\epsf@fileline:. \\%
       \fi
   \ifnot@eof\repeat
   }\closein\ps@stream\fi}%
%
%
\long\def\epsf@test#1#2#3:#4\\{\def\epsf@testit{#1#2}
			\ifx\epsf@testit\epsf@start\else
\ps@typeout{Warning! File does not start with `\epsf@start'.  It may not be a
PostScript file.}
			\fi
			\@psfiletrue} 
%
%
{\catcode`\%=12\global\let\epsf@percent=
%
%
%
\long\def\epsf@aux#1#2:#3\\{\ifx#1\epsf@percent
   \def\epsf@testit{#2}\ifx\epsf@testit\epsf@bblit
	\@atendfalse
        \epsf@atend #3 . \\%
	\if@atend
	   \if@verbose{
		\ps@typeout{psfig: found `(atend)'; continuing search}
	   }\fi
        \else
        \epsf@grab #3 . . . \\%
        \not@eoffalse
        \global\no@bbfalse
        \fi
   \fi\fi}%
%
%
\def\epsf@grab #1 #2 #3 #4 #5\\{%
   \global\def\epsf@llx{#1}\ifx\epsf@llx\empty
      \epsf@grab #2 #3 #4 #5 .\\\else
   \global\def\epsf@lly{#2}%
   \global\def\epsf@urx{#3}\global\def\epsf@ury{#4}\fi}%
%
%
\def\epsf@atendlit{(atend)}
\def\epsf@atend #1 #2 #3\\{%
   \def\epsf@tmp{#1}\ifx\epsf@tmp\empty
      \epsf@atend #2 #3 .\\\else
   \ifx\epsf@tmp\epsf@atendlit\@atendtrue\fi\fi}


\chardef\psletter = 11 
\chardef\other = 12

\newif \ifdebug 
\newif\ifc@mpute 
\c@mputetrue 

\let\then = \relax
\def\r@dian{pt }
\let\r@dians = \r@dian
\let\dimensionless@nit = \r@dian
\let\dimensionless@nits = \dimensionless@nit
\def\internal@nit{sp }
\let\internal@nits = \internal@nit
\newif\ifstillc@nverging
\def \Mess@ge #1{\ifdebug \then \message {#1} \fi}

{ 
	\catcode `\@ = \psletter
	\gdef \nodimen {\expandafter \n@dimen \the \dimen}
	\gdef \term #1 #2 #3%
	       {\edef \t@ {\the #1}
		\edef \t@@ {\expandafter \n@dimen \the #2\r@dian}%
		\t@rm {\t@} {\t@@} {#3}%
	       }
	\gdef \t@rm #1 #2 #3%
	       {{%
		\count 0 = 0
		\dimen 0 = 1 \dimensionless@nit
		\dimen 2 = #2\relax
		\Mess@ge {Calculating term #1 of \nodimen 2}%
		\loop
		\ifnum	\count 0 < #1
		\then	\advance \count 0 by 1
			\Mess@ge {Iteration \the \count 0 \space}%
			\Multiply \dimen 0 by {\dimen 2}%
			\Mess@ge {After multiplication, term = \nodimen 0}%
			\Divide \dimen 0 by {\count 0}%
			\Mess@ge {After division, term = \nodimen 0}%
		\repeat
		\Mess@ge {Final value for term #1 of
				\nodimen 2 \space is \nodimen 0}%
		\xdef \Term {#3 = \nodimen 0 \r@dians}%
		\aftergroup \Term
	       }}
	\catcode `\p = \other
	\catcode `\t = \other
	\gdef \n@dimen #1pt{#1} 
}

\def \Divide #1by #2{\divide #1 by #2} 

\def \Multiply #1by #2
       {{
	\count 0 = #1\relax
	\count 2 = #2\relax
	\count 4 = 65536
	\Mess@ge {Before scaling, count 0 = \the \count 0 \space and
			count 2 = \the \count 2}%
	\ifnum	\count 0 > 32767 
	\then	\divide \count 0 by 4
		\divide \count 4 by 4
	\else	\ifnum	\count 0 < -32767
		\then	\divide \count 0 by 4
			\divide \count 4 by 4
		\else
		\fi
	\fi
	\ifnum	\count 2 > 32767 
	\then	\divide \count 2 by 4
		\divide \count 4 by 4
	\else	\ifnum	\count 2 < -32767
		\then	\divide \count 2 by 4
			\divide \count 4 by 4
		\else
		\fi
	\fi
	\multiply \count 0 by \count 2
	\divide \count 0 by \count 4
	\xdef \product {#1 = \the \count 0 \internal@nits}%
	\aftergroup \product
       }}

\def\r@duce{\ifdim\dimen0 > 90\r@dian \then   
		\multiply\dimen0 by -1
		\advance\dimen0 by 180\r@dian
		\r@duce
	    \else \ifdim\dimen0 < -90\r@dian \then  
		\advance\dimen0 by 360\r@dian
		\r@duce
		\fi
	    \fi}

\def\Sine#1%
       {{%
	\dimen 0 = #1 \r@dian
	\r@duce
	\ifdim\dimen0 = -90\r@dian \then
	   \dimen4 = -1\r@dian
	   \c@mputefalse
	\fi
	\ifdim\dimen0 = 90\r@dian \then
	   \dimen4 = 1\r@dian
	   \c@mputefalse
	\fi
	\ifdim\dimen0 = 0\r@dian \then
	   \dimen4 = 0\r@dian
	   \c@mputefalse
	\fi
	\ifc@mpute \then
		\divide\dimen0 by 180
		\dimen0=3.141592654\dimen0
		\dimen 2 = 3.1415926535897963\r@dian 
		\divide\dimen 2 by 2 
		\Mess@ge {Sin: calculating Sin of \nodimen 0}%
		\count 0 = 1 
		\dimen 2 = 1 \r@dian 
		\dimen 4 = 0 \r@dian 
		\loop
			\ifnum	\dimen 2 = 0 
			\then	\stillc@nvergingfalse
			\else	\stillc@nvergingtrue
			\fi
			\ifstillc@nverging 
			\then	\term {\count 0} {\dimen 0} {\dimen 2}%
				\advance \count 0 by 2
				\count 2 = \count 0
				\divide \count 2 by 2
				\ifodd	\count 2 
				\then	\advance \dimen 4 by \dimen 2
				\else	\advance \dimen 4 by -\dimen 2
				\fi
		\repeat
	\fi
			\xdef \sine {\nodimen 4}%
       }}

\def\Cosine#1{\ifx\sine\UnDefined\edef\Savesine{\relax}\else
		             \edef\Savesine{\sine}\fi
	{\dimen0=#1\r@dian\advance\dimen0 by 90\r@dian
	 \Sine{\nodimen 0}
	 \xdef\cosine{\sine}
	 \xdef\sine{\Savesine}}}

\def\psdraft{
	\def\@psdraft{0}
}
\def\psfull{
	\def\@psdraft{100}
}

\psfull

\newif\if@scalefirst
\def\psscalefirst{\@scalefirsttrue}
\def\psrotatefirst{\@scalefirstfalse}
\psrotatefirst

\newif\if@draftbox
\def\psnodraftbox{
	\@draftboxfalse
}
\def\psdraftbox{
	\@draftboxtrue
}
\@draftboxtrue

\newif\if@prologfile
\newif\if@postlogfile
\def\pssilent{
	\@noisyfalse
}
\def\psnoisy{
	\@noisytrue
}
\psnoisy
\newif\if@bbllx
\newif\if@bblly
\newif\if@bburx
\newif\if@bbury
\newif\if@height
\newif\if@width
\newif\if@rheight
\newif\if@rwidth
\newif\if@angle
\newif\if@clip
\newif\if@verbose
\def\@p@@sclip#1{\@cliptrue}
\newif\if@decmpr
\def\@p@@sfigure#1{\def\@p@sfile{null}\def\@p@sbbfile{null}\@decmprfalse
   \openin1=\ps@predir#1
   \ifeof1
	\closein1
	\get@dir{#1}
	\ifx\ps@founddir\leer
		\openin1=\ps@predir#1.bb
		\ifeof1
			\closein1
			\get@dir{#1.bb}
			\ifx\ps@founddir\leer
				\ps@typeout{Can't find #1 in \figurepath}
			\else
				\@decmprtrue
				\def\@p@sfile{\ps@founddir\ps@dir#1}
				\def\@p@sbbfile{\ps@founddir\ps@dir#1.bb}
			\fi
		\else
			\closein1
			\@decmprtrue
			\def\@p@sfile{#1}
			\def\@p@sbbfile{#1.bb}
		\fi
	\else
		\def\@p@sfile{\ps@founddir\ps@dir#1}
		\def\@p@sbbfile{\ps@founddir\ps@dir#1}
	\fi
   \else
	\closein1
	\def\@p@sfile{#1}
	\def\@p@sbbfile{#1}
   \fi
}
\def\@p@@sfile#1{\@p@@sfigure{#1}}
\def\@p@@sbbllx#1{
		\@bbllxtrue
		\dimen100=#1
		\edef\@p@sbbllx{\number\dimen100}
}
\def\@p@@sbblly#1{
		\@bbllytrue
		\dimen100=#1
		\edef\@p@sbblly{\number\dimen100}
}
\def\@p@@sbburx#1{
		\@bburxtrue
		\dimen100=#1
		\edef\@p@sbburx{\number\dimen100}
}
\def\@p@@sbbury#1{
		\@bburytrue
		\dimen100=#1
		\edef\@p@sbbury{\number\dimen100}
}
\def\@p@@sheight#1{
		\@heighttrue
		\dimen100=#1
   		\edef\@p@sheight{\number\dimen100}
}
\def\@p@@swidth#1{
		\@widthtrue
		\dimen100=#1
		\edef\@p@swidth{\number\dimen100}
}
\def\@p@@srheight#1{
		\@rheighttrue
		\dimen100=#1
		\edef\@p@srheight{\number\dimen100}
}
\def\@p@@srwidth#1{
		\@rwidthtrue
		\dimen100=#1
		\edef\@p@srwidth{\number\dimen100}
}
\def\@p@@sangle#1{
		\@angletrue
		\edef\@p@sangle{#1} 
}
\def\@p@@ssilent#1{
		\@verbosefalse
}
\def\@p@@sprolog#1{\@prologfiletrue\def\@prologfileval{#1}}
\def\@p@@spostlog#1{\@postlogfiletrue\def\@postlogfileval{#1}}
\def\@cs@name#1{\csname #1\endcsname}
\def\@setparms#1=#2,{\@cs@name{@p@@s#1}{#2}}
%
%
\def\ps@init@parms{
		\@bbllxfalse \@bbllyfalse
		\@bburxfalse \@bburyfalse
		\@heightfalse \@widthfalse
		\@rheightfalse \@rwidthfalse
		\def\@p@sbbllx{}\def\@p@sbblly{}
		\def\@p@sbburx{}\def\@p@sbbury{}
		\def\@p@sheight{}\def\@p@swidth{}
		\def\@p@srheight{}\def\@p@srwidth{}
		\def\@p@sangle{0}
		\def\@p@sfile{} \def\@p@sbbfile{}
		\def\@p@scost{10}
		\def\@sc{}
		\@prologfilefalse
		\@postlogfilefalse
		\@clipfalse
		\if@noisy
			\@verbosetrue
		\else
			\@verbosefalse
		\fi
}
%
%
\def\parse@ps@parms#1{
	 	\@psdo\@psfiga:=#1\do
		   {\expandafter\@setparms\@psfiga,}}
%
%
\newif\ifno@bb
\def\bb@missing{
	\if@verbose{
		\ps@typeout{psfig: searching \@p@sbbfile \space  for bounding box}
	}\fi
	\no@bbtrue
	\epsf@getbb{\@p@sbbfile}
        \ifno@bb \else \bb@cull\epsf@llx\epsf@lly\epsf@urx\epsf@ury\fi
}
\def\bb@cull#1#2#3#4{
	\dimen100=#1 bp\edef\@p@sbbllx{\number\dimen100}
	\dimen100=#2 bp\edef\@p@sbblly{\number\dimen100}
	\dimen100=#3 bp\edef\@p@sbburx{\number\dimen100}
	\dimen100=#4 bp\edef\@p@sbbury{\number\dimen100}
	\no@bbfalse
}
\newdimen\p@intvaluex
\newdimen\p@intvaluey
\def\rotate@#1#2{{\dimen0=#1 sp\dimen1=#2 sp
		  \global\p@intvaluex=\cosine\dimen0
		  \dimen3=\sine\dimen1
		  \global\advance\p@intvaluex by -\dimen3
		  \global\p@intvaluey=\sine\dimen0
		  \dimen3=\cosine\dimen1
		  \global\advance\p@intvaluey by \dimen3
		  }}
\def\compute@bb{
		\no@bbfalse
		\if@bbllx \else \no@bbtrue \fi
		\if@bblly \else \no@bbtrue \fi
		\if@bburx \else \no@bbtrue \fi
		\if@bbury \else \no@bbtrue \fi
		\ifno@bb \bb@missing \fi
		\ifno@bb \ps@typeout{FATAL ERROR: no bb supplied or found}
			\no-bb-error
		\fi
		%
%
		\count203=\@p@sbburx
		\count204=\@p@sbbury
		\advance\count203 by -\@p@sbbllx
		\advance\count204 by -\@p@sbblly
		\edef\ps@bbw{\number\count203}
		\edef\ps@bbh{\number\count204}
		\if@angle
			\Sine{\@p@sangle}\Cosine{\@p@sangle}
	        	{\dimen100=\maxdimen\xdef\r@p@sbbllx{\number\dimen100}
					    \xdef\r@p@sbblly{\number\dimen100}
			                    \xdef\r@p@sbburx{-\number\dimen100}
					    \xdef\r@p@sbbury{-\number\dimen100}}
%
                        \def\minmaxtest{
			   \ifnum\number\p@intvaluex<\r@p@sbbllx
			      \xdef\r@p@sbbllx{\number\p@intvaluex}\fi
			   \ifnum\number\p@intvaluex>\r@p@sbburx
			      \xdef\r@p@sbburx{\number\p@intvaluex}\fi
			   \ifnum\number\p@intvaluey<\r@p@sbblly
			      \xdef\r@p@sbblly{\number\p@intvaluey}\fi
			   \ifnum\number\p@intvaluey>\r@p@sbbury
			      \xdef\r@p@sbbury{\number\p@intvaluey}\fi
			   }
			\rotate@{\@p@sbbllx}{\@p@sbblly}
			\minmaxtest
			\rotate@{\@p@sbbllx}{\@p@sbbury}
			\minmaxtest
			\rotate@{\@p@sbburx}{\@p@sbblly}
			\minmaxtest
			\rotate@{\@p@sbburx}{\@p@sbbury}
			\minmaxtest
			\edef\@p@sbbllx{\r@p@sbbllx}\edef\@p@sbblly{\r@p@sbblly}
			\edef\@p@sbburx{\r@p@sbburx}\edef\@p@sbbury{\r@p@sbbury}
		\fi
		\count203=\@p@sbburx
		\count204=\@p@sbbury
		\advance\count203 by -\@p@sbbllx
		\advance\count204 by -\@p@sbblly
		\edef\@bbw{\number\count203}
		\edef\@bbh{\number\count204}
}
%
%
\def\in@hundreds#1#2#3{\count240=#2 \count241=#3
		     \count100=\count240	
		     \divide\count100 by \count241
		     \count101=\count100
		     \multiply\count101 by \count241
		     \advance\count240 by -\count101
		     \multiply\count240 by 10
		     \count101=\count240	
		     \divide\count101 by \count241
		     \count102=\count101
		     \multiply\count102 by \count241
		     \advance\count240 by -\count102
		     \multiply\count240 by 10
		     \count102=\count240	
		     \divide\count102 by \count241
		     \count200=#1\count205=0
		     \count201=\count200
			\multiply\count201 by \count100
		 	\advance\count205 by \count201
		     \count201=\count200
			\divide\count201 by 10
			\multiply\count201 by \count101
			\advance\count205 by \count201
		     \count201=\count200
			\divide\count201 by 100
			\multiply\count201 by \count102
			\advance\count205 by \count201
		     \edef\@result{\number\count205}
}
\def\compute@wfromh{
		\in@hundreds{\@p@sheight}{\@bbw}{\@bbh}
		\edef\@p@swidth{\@result}
}
\def\compute@hfromw{
	        \in@hundreds{\@p@swidth}{\@bbh}{\@bbw}
		\edef\@p@sheight{\@result}
}
\def\compute@handw{
		\if@height
			\if@width
			\else
				\compute@wfromh
			\fi
		\else
			\if@width
				\compute@hfromw
			\else
				\edef\@p@sheight{\@bbh}
				\edef\@p@swidth{\@bbw}
			\fi
		\fi
}
\def\compute@resv{
		\if@rheight \else \edef\@p@srheight{\@p@sheight} \fi
		\if@rwidth \else \edef\@p@srwidth{\@p@swidth} \fi
}
%
\def\compute@sizes{
	\compute@bb
	\if@scalefirst\if@angle
	\if@width
	   \in@hundreds{\@p@swidth}{\@bbw}{\ps@bbw}
	   \edef\@p@swidth{\@result}
	\fi
	\if@height
	   \in@hundreds{\@p@sheight}{\@bbh}{\ps@bbh}
	   \edef\@p@sheight{\@result}
	\fi
	\fi\fi
	\compute@handw
	\compute@resv}
\def\OzTeXSpecials{
	\special{empty.ps /@isp {true} def}
	\special{empty.ps \@p@swidth \space \@p@sheight \space
			\@p@sbbllx \space \@p@sbblly \space
			\@p@sbburx \space \@p@sbbury \space
			startTexFig \space }
	\if@clip{
		\if@verbose{
			\ps@typeout{(clip)}
		}\fi
		\special{empty.ps doclip \space }
	}\fi
	\if@angle{
		\if@verbose{
			\ps@typeout{(rotate)}
		}\fi
		\special {empty.ps \@p@sangle \space rotate \space}
	}\fi
	\if@prologfile
	    \special{\@prologfileval \space } \fi
	\if@decmpr{
		\if@verbose{
			\ps@typeout{psfig: Compression not available
			in OzTeX version \space }
		}\fi
	}\else{
		\if@verbose{
			\ps@typeout{psfig: including \@p@sfile \space }
		}\fi
		\special{epsf=\ps@predir\@p@sfile \space }
	}\fi
	\if@postlogfile
	    \special{\@postlogfileval \space } \fi
	\special{empty.ps /@isp {false} def}
}
\def\DvipsSpecials{
	\special{ps::[begin] 	\@p@swidth \space \@p@sheight \space
			\@p@sbbllx \space \@p@sbblly \space
			\@p@sbburx \space \@p@sbbury \space
			startTexFig \space }
	\if@clip{
		\if@verbose{
			\ps@typeout{(clip)}
		}\fi
		\special{ps:: doclip \space }
	}\fi
	\if@angle
		\if@verbose{
			\ps@typeout{(clip)}
		}\fi
		\special {ps:: \@p@sangle \space rotate \space}
	\fi
	\if@prologfile
	    \special{ps: plotfile \@prologfileval \space } \fi
	\if@decmpr{
		\if@verbose{
			\ps@typeout{psfig: including \@p@sfile.Z \space }
		}\fi
		\special{ps: plotfile "`zcat \@p@sfile.Z" \space }
	}\else{
		\if@verbose{
			\ps@typeout{psfig: including \@p@sfile \space }
		}\fi
		\special{ps: plotfile \@p@sfile \space }
	}\fi
	\if@postlogfile
	    \special{ps: plotfile \@postlogfileval \space } \fi
	\special{ps::[end] endTexFig \space }
}
%
%
\def\psfig#1{\vbox {
	%
	\ps@init@parms
	\parse@ps@parms{#1}
	\compute@sizes
	\ifnum\@p@scost<\@psdraft{
		\PsfigSpecials
		\vbox to \@p@srheight sp{
			\hbox to \@p@srwidth sp{
				\hss
			}
		\vss
		}
	}\else{
		\if@draftbox{
			\hbox{\fbox{\vbox to \@p@srheight sp{
			\vss
			\hbox to \@p@srwidth sp{ \hss
			 \hss }
			\vss
			}}}
		}\else{
			\vbox to \@p@srheight sp{
			\vss
			\hbox to \@p@srwidth sp{\hss}
			\vss
			}
		}\fi

	}\fi
}}
\psfigRestoreAt
\setDriver
\let\@=\LaTeXAtSign

\overfullrule=0pt
\baselineskip=15truept
%
%

\def\MPA#1#2{Max-Planck-Institut f\"ur Astrophysik 19{#1},
             Preprint {$\underline{#2}$} }
 \def\z{\phantom 1}
 \def\Hbar{$\overline H\ $}
 \def\Htil{$\widetilde H\ $}
 \def\Etilbv{$\widetilde E_{B-V}\ $}
 \def\etal{{et al.} \thinspace}
 \def\eg{{e.g.,} \thinspace}
 \def\ie{{i.e.,} \thinspace}
 \def\eck#1{\left\lbrack #1 \right\rbrack}
 \def\eqck#1{$\bigl\lbrack$ #1 $\bigr\rbrack$}
 \def\rund#1{\left( #1 \right)}
 \def\ave#1{\langle #1 \rangle}
 \def\:{\mskip\medmuskip}                         
 \def\lb{\lbrack} \def\rb{\rbrack}                
 \def\unit#1{\nobreak{\:{\rm#1}}}                 
 \def\inunits#1{\nobreak{\:\lb{\rm#1}\rb}}        
 \def\gcc{gcm$^{-3}$}
 \def\mstar{ M_{\ast} }
 \def\msol{ M_\odot }            
 \def\ms{ M_\odot }              
 \def\lsol{ L_\odot }            
 \def\ni{$^{56}{\rm Ni}\ $}      
 \def\Ni{$^{56}{\rm Ni}$}      
 \def\co{$^{56}{\rm Co}\ $}      
 \def\Co{$^{56}{\rm Co}$}      
 \def\fe{$^{56}{\rm Fe}\ $}      
 \def\kelvin{\thinspace\rm{\sp{o}{\kern-.08333em }K}\ }
%
\nopagenumbers
\title{Hard X- and Gamma-Rays from Type Ia Supernovae}
\bigskip
\bigskip
\centerline{P.~H\"oflich$^{1,2}$, J.C. Wheeler $^1$, A. Khokhlov $^3$}
\bigskip
\leftline{1. Department of Astronomy, University of Texas, Austin, TX 07871,
USA}
\leftline{2. Department of Physics, University of Basel, CH-7046 Basel,
Switzerland}
\leftline{3. Lab. for Computational Physics and Fluid Dynamics,}
\leftline{ Code 6404,  NRL, Washington, DC 20375, USA}
\bigskip

\heading{Abstract}

\noindent
 The $\gamma$-ray light curves and spectra are presented for a set of
theoretical Type~Ia supernova models including deflagration, detonation,
delayed detonation, and pulsating delayed detonations of Chandrasekhar mass
white dwarfs as well as merger scenarios that may involve more than
the Chandrasekhar mass and helium detonations of
sub-Chandrasekhar mass white dwarfs.
 The results have been obtained with a
Monte Carlo radiation transport scheme which takes into account all
relevant $\gamma$-transitions and interaction processes.
 The result is a set of accurate line profiles which
are characteristic of the initial \Ni-mass
distribution of the supernova models.
 The $\gamma$-rays probe the isotopic rather than just the elemental
distribution of the radioactive elements in the ejecta.
 Details of the line profiles including the line width, shift with respect
to the rest frame, and line ratios are discussed.
 With sufficient energy and temporal resolution,
different model scenarios can clearly be distinguished.
 Observational strategies are discussed for current and immediately upcoming
generations of satellites (CGRO and INTEGRAL) as well as
projected future missions including concepts such as Laue telescopes.
 With CGRO, it is currently possible with sufficiently early
observations (near optical maximum) to distinguish helium detonations
from explosions of Chandrasekhar mass progenitors and
of those involving mergers up to a distance of about 15 Mpc.
This translates into one target of opportunity every eight years.
 SNe~Ia up to about 10 Mpc would allow detailed CGRO studies of line ratios
of \co lines.
 INTEGRAL will be able to perform detailed studies
of the \co line profiles with a range comparable to CGRO.
 The superior sensitivity of INTEGRAL for low energies
makes detection and detailed study of the positron annhilation
line and appropriate low energy \ni lines possible up to about 10 - 15 Mpc
for all models.  This capability means that this lower energy range may be
the most useful for INTEGRAL detection and study of SNe~Ia.
 Such studies will allow the determination of the precise time of
the explosion.
 Whereas the current generation of $\gamma$-ray detectors will allow the study
of supernovae which are discovered by other means, a new generation
of proposed $\gamma$-ray detectors with sensitivity of
about $10^{-6}$ photons~sec$^{-1}$~cm$^{-2}$ would generate the opportunity to
discover supernovae by their $\gamma$-ray emission
up to a distance of $\approx $ 100 Mpc.
 This would allow a systematic study of the variety of SNe~Ia in terms of
their $\gamma$-ray properties, independent of their optical properties.
 In addition, since $\gamma $-rays are not obscured by the host galaxy,
such experiments would, for the first time, provide absolute supernova
rates.
 Relative rates as a function of the morphology of and position in
the host galaxy could be studied directly.

\bigskip \noindent
{\it Subject headings:} Supernovae and supernovae remnants: general --
gamma rays

%
%
%
\section{Introduction}

 Type Ia Supernovae (SNe~Ia) are among the most spectacular explosive
events.
 They are major contributors to the production
of heavy elements and hence a critical component for understanding the
life cycles of matter in the Universe and the chemical evolution of galaxies.
 While much progress has been made on understanding the underlying physics
of the thermonuclear explosion (Khokhlov, Oran and Wheeler 1997ab;
 Niemeyer \& Woosley 1997), the basic processes require further
elucidation and testing and the progenitor evolution is still a
deep mystery (Wheeler 1996).
 The great brightness of SNe~Ia has made them a valuable tool for
the measurement of extragalactic distances (Freedman  et al. 1994)
and to determine the shape of the Universe (Perlmutter et al. 1995,
Schmidt et al. 1997).
 The absolute brightness must be known either
by using other distance calibrators (Hamuy et al. 1996,
Ries, Press, \& Kirshner 1996, Sandage \& Tammann 1996)
 or theoretical models for the light curves
and spectra (H\"oflich 1995, H\"oflich \& Khokhlov 1996, HK96 hereafter).
 A detailed physical understanding of SNe~Ia is important for
its own sake.
 In addition, such a physical understanding is necessary to constrain
systematic errors when using them at high redshifts to determine the
deceleration parameter $q_o$ and other cosmological parameters.
 Evolutionary effects on the progenitor population, for instance
the initial IMF and the metallicity, are expected to produce
systematic errors (H\"oflich et al. 1997).

 It is widely accepted that SNe~Ia are thermonuclear explosions of
carbon-oxygen white dwarfs; however, rather different progenitor
models are under discussion.   Three main scenarios can be distinguished.
The first group consists of carbon-oxygen white
dwarfs with a mass close to the Chandrasekhar mass which accrete mass
through Roche-lobe overflow from an an unevolved (CV-like) or
evolved companion star.
 In these accretion models, the explosion is
triggered by compressional heating.
 From the theoretical standpoint, the key questions are
the condition under which the thermonuclear explosion starts
and how the flame propagates through the white dwarf.  The ignition
and propagation may be affected by the rotation and mass accretion
rate of the white dwarf which will in turn be affected by the
metallicity, age, and population of the progenitor system.
 The second group of progenitor
models consists of two low-mass white dwarfs in a close orbit which
decays due to the emission of gravitational radiation.  This eventually
leads to the merging of the two white dwarfs.  In this scenario, the total
mass of the merged object may exceed the Chandrasekhar mass and
rotation may play an especially significant role.
      A third class of models are those involving helium detonations,
i.e. double detonation of a C/O-white dwarf triggered by the detonation of
an outer helium layer in low-mass white dwarfs.

 Each of these scenarios will have different implications for
the resulting supernova statistics, the evolutionary ages of
the progenitor systems and for chemical evolution of galaxies and
the Universe.

 Gamma rays are of particular interest as a diagnostic tool of these
various progenitor models because they allow the direct observation
of radioactive isotopes which power the observable light curves
and spectra.
 Almost three decades ago Clayton et al. (1969) pointed out that
the $\gamma$-ray lines from SN~Ia explosions should be detectable at distances
up to a few Mpc. Since then many authors have studied the gamma-ray
signature of SN~Ia (Clayton 1974; Ambwani \& Sutherland 1988; Chan \&
Lingenfelter 1988, 1990, 1991; Burrows \& The 1990;
Burrows \etal 1991; H\"oflich, Khokhlov \& M\"uller
 1992, 1993ab,  1994, Kumagai \& Nomoto 1997).
 These studies have shown that $\gamma$-ray observations of SNe~Ia
provide a powerful diagnostic tool to test proposed theoretical
models.
 The key role of $\gamma$-rays for understanding of SNe~Ia became obvious
in the case of the peculiar SN 1991T.
 Because of the lack of $\gamma$-ray detection and/or firm upper limits,
the nature of SN1991T was not fully revealed despite various efforts
(H\"oflich et al., 1994, Kumagai \& Nomoto 1997) and an excellent set
of data from the optical to the IR (e.g. Meikle et al. 1996)
and even a distance from $\delta$-Cephei stars (Saha et al. 1996).
 Nevertheless, this event clearly showed the potential and implications
for $\gamma$-ray observations (e.g. Lichti et al. 1994;
Leising et al. 1995, and references therein).

 Although  systematic studies of the $\gamma$-ray emission properties
of different scenarios have been published, several scenarios, some
of them with rather strong observational and physical support,
have not been considered (e.g. Burrows \& The 1990; Chan \& Lingenfelter 1990,
1991;
H\"oflich et al. 1992) or where both Chandrasekar and
sub-Chandrasekar type models have been included, no
detailed study of line profiles has been performed (e.g.  Kumagai \& Nomoto
1997).

 In addition, little attention has been paid to specific observational
strategies and the information required to allow
instrument designers to estimate the performance of their instruments
and to analyse future data.
 A detailed knowledge of the line position is critical
for instruments which provide only a very limited wavelength range
such as the Laue telescopes currently in the design phase.
 Even for conventional $\gamma$-ray telescopes such as CGRO or the upcoming
INTEGRAL, the detection limits may improve if information on the
line position and profile is taken into account. Due to the low signal
to noise ratio, the detectibility depends sensitively on the line
profile and on the knowledge of the line position.
 In principle, the information from the models and the measurement
can be combined during the data analysis to filter out the high noise
 rather than comparing predicted  and observed fluxes after the
reductions.

 This paper will seek to fill some of the gaps just mentioned.
 In the next section we describe briefly the numerical methods and
the hydrodynamical models on which this study is based.
 In \S~3 the properties of line profiles are quantitatively discussed
along with the $\gamma$-ray light curves.
 In \S~4, the expected rate and the possible applications for
supernova searches with current and future instruments are
briefly discussed.
 \S~5 presents a final discussion and conclusions.

\section{Brief Description of the Numerical Methods}

\subsection { Hydrodynamics}

 The dynamical explosion models are calculated using a one-dimensional
Lagrangian hydro code with an artificial viscosity (Khokhlov, 1991)
and a radiation-hydro code that includes nuclear networks
(HK96 and references therein).
 The latter code  is based on the light curve code
presented in (H\"oflich, M\"uller \& Khokhlov 1993a)
that solves the hydrodynamical equations explicitly
by the piecewise parabolic method (Collela and Woodward 1984).
This code includes the solution of the radiation transport
implicitly via the moment equations, expansion opacities,
and a  detailed equation of state.
 Typically,  300 to  500 depth points are used.
 The $\gamma$-ray transport is omitted during the hydrodynamical phase
because of the high optical depth of the inner layers.
 Nuclear burning is taken into account using Thielemann's network
(Thielemann, Nomoto \& Hashimoto 1994 and references therein).
 During the hydro, an $\alpha$-network of 20 isotopes is considered
to properly describe the energy release.
 The final chemical structure is calculated  by postprocessing the
hydrodynamical model using a network of 216 isotopes.

\subsection{Gamma-ray transport}

For a given explosion model, $\gamma$-rays are calculated using a detailed
Monte Carlo $\gamma$-ray transport scheme (H\"oflich et al. 1992, 1993b).
 A Monte Carlo scheme has been used because of its flexibility with respect to
the treatment of physical processes, because relativistic effects can
be included in a straightforward way, and because it allows for
a consistent treatment of both the continua and lines.

Photons are initialized in the transport code according to
the distribution of \ni and \co in the hydrodynamical model.
All \ni (6) and  \co (46) decay lines
have been taken into account. For the $\beta ^+$ decay, the assumption
is made that all positrons form positronium at the point where they are
produced by radioactive decay and hence that they decay locally to produce
annihilation photons.
 The positron leakage can be neglected during the first few months,
but it may become important during late stages, especially for low mass
envelopes or if the \ni is located in the outer layers.
 For a discussion of positron leakage, see Colgate, Fryer \& Hand (1997)
and references therein.
 About $1-5~10^6$ initial decay photons are created for
each Monte Carlo calculation.
 The direction of the photons is chosen
randomly in the comoving frame of the expanding matter.
Following Ambwani \& Sutherland (1988), we have included the following
three interaction processes in the Monte Carlo transport code:  Compton
scattering according to the full angle-dependent
Klein-Nishina formula, pair-production with cross section
taken from Hubbell (1969),
and photoelectric absorption with cross sections
determined on the basis of the tables given by Veigle (1973).
 The emergent $\gamma$-ray spectra have been calculated with a high
energy resolution of $\lambda/\delta \lambda = 600 $.
 This resolution is sufficient to obtain accurate line profiles and fluxes.
For more details see H\"oflich \etal (1992).

\section{Dynamical Models}

 The $\gamma$-ray spectra have been computed for a large variety
of models including delayed detonations (the DD-series
of H\"oflich et al. 1997), pulsating delayed detonations (the
PDD-series presented in  Khokhlov 1992, H\"oflich, Khokhlov \& Wheeler, 1995,
and HK96), binary white dwarf
merger scenarios (the DET2env series of Khokhlov, M\"uller \& H\"oflich
1993),
and helium-detonation models (the HeD series of HK96).
The classical detonation scenarios are omitted because this class of
models is known to produce hardly any intermediate mass elements and
hence to fail by inspection the criterion of agreement
with spectral observations.

 The first group of models we consider (the delayed detonation
and pulsating delayed detonation models)
consists of carbon-oxygen white dwarfs with a
mass close to the Chandrasekhar limit.  These systems presumably
result from  accretion through Roche-lobe overflow from a companion
star (Nomoto \& Sugimoto 1977; Nomoto 1982).  In these accretion models,
the explosion is triggered by compressional heating and the heat
wave traveling inwards from any nuclear shell-burning region at the
surface. Typically, the thermonuclear runaway occurs
at central densities $\rho _c \approx 2 - 3\times10^9$ g~cm$^{-3}$.
In the innermost regions, burning occurs at low $Y_e$ and,
consequently, neutron-rich isotopes are produced rather
than \ni as shown in Fig. 1 (panel 1).

 \begfig -.2cm
 \psfig{figure=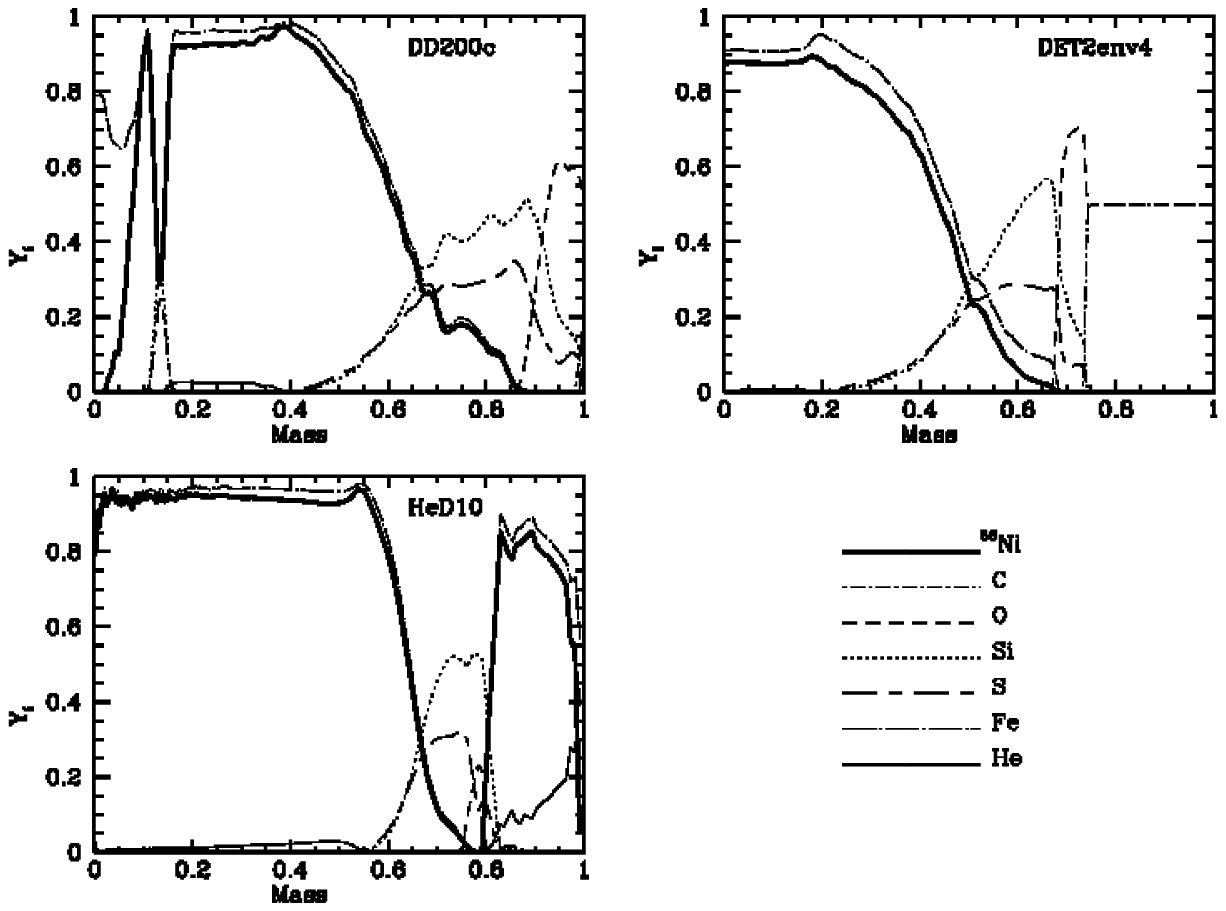,width=13.6cm,rwidth=14.5cm,clip=,angle=270}
\figure{1}
{Some abundances and $^{56} Ni$ distribution as a function of mass in stellar
units for
the delayed detonation model DD200c, the merger model DET2env4 and the helium
detonation
model HeD10.}
\endfig

 The second group of progenitor
models is based on the scenario of two low-mass white dwarfs in
a close orbit which decays due to the emission of gravitational
radiation thus leading eventually to the merging of the two
white dwarfs as the smaller mass, larger radius white dwarf
fills its Roche lobe first and undergoes unstable mass transfer
(Iben and Tutukov 1984, Webbink 1984, Iben and Tutukov 1997).
The combined mass of the two white dwarfs exceeds the Chandrasekhar limit,
but the larger mass white dwarf which receives the mass is nevertheless
itself less than the Chandrasekhar limit.  Its subsequent
carbon ignition is expected to occur at low density as mass
accretes from the extended C/O envelope formed by the disrupted star
(Benz et al. 1989).  Since burning occurs at low densities,
there is little neutronization and \ni is produced in the center (Fig. 1, panel
2).

 The third class of models explored here are based on
low mass white dwarfs in which a degenerate helium layer builds
up on top of a degenerate C/O core.  The explosion is
triggered by detonation of the helium layer which then, in
some circumstances, causes the inner C/O core to detonate as well.
This scenario for normally bright SNe~Ia was explored by Nomoto (1980),
Woosley, Weaver \& Taam (1980), and most recently by Livne \& Glasner (1990),
Woosley \& Weaver (1994), Livne \& Arnett (1995),
and HK96.
This scenario has also been suggested as an explanation of
the subluminous SNe~Ia (Woosley \& Weaver 1994, Ruiz-Lapuente et al. 1993).
 In this class of models, the thermonuclear runaway starts
near the bottom of the outer He layer.
 The \ni distribution, and the subsequent $\gamma$-ray emission
differs significantly from the models based on the other scenarios.
Due to the large energy release during He burning,
$\approx 0.1 M_\odot$ of \ni is produced in the outer region
originally composed of helium.
 Subsequently, a detonation travels inwards and ignites
the C/O core at low density, yielding a central concentration of
\ni at lower velocity.  This bi-modal \ni distributions differs
significantly from the that expected in the other scenarios (Fig.1, panel 3).
Depending on model parameters, the amount of this inner \ni
can be substantial (see model HeD10 in Table 1) and hence represent
a potential normally bright SN~Ia or it can be as small as the
outer layer, $\approx 0.1 M_\odot$ (see model HeD6 in Table 1),
which could represent a subluminous event.

\begtabsmall
\table{1}{ Parameters of Dynamical Models of SNe~Ia.  }

\halign{#\hfil&&\quad#\hfil\cr}
\hline
\+Model~~~~~~~~~ & Mode of~~~~~~~~ & $M_\star$ ~~~~~~~~~& $\rho_c$ ~~~~~~~~~ &
$\alpha$ ~~~~~~~~~&
  $\rho_{tr}$~~~~~~~~~ &
                       $M_{Ni}$~~~~~~~~                  \cr
\+& explosion & $[ M_\odot ]$ & $[10^9]$ &  &   $[ 10^7]$ &      $[ M_\odot ]$
     \cr
\hline
\hline
\+DD200c  & delayed det. &  1.4  &  2.0   &  0.03  &  2.0  &
                           0.613               \cr
\+DD201c  & delayed det. &  1.4  &  2.0   &  0.03  &  1.7  &
                           0.56                \cr
\+DD202c  & delayed det. &  1.4  &  2.0   &  0.03  &  2.5  &
                           0.67                \cr
\+DD203c (2:3)& delayed det. &  1.4  &  2.0   &  0.03  &  2.0  &
                           0.59                \cr
\hline
\+PDD5    & pul.del.det. &  1.4  &  2.7   &  0.03  &  0.76 &
                           0.12                       \cr
\+PDD6    & pul.del.det. &  1.4  &  2.7   &  0.03  &  2.2  &
                           0.56                         \cr
\hline
\+HeD6    & He-det.      &  0.6+0.172 &  .0091  &  ---   &  ---  &
                           0.252 (.08)             \cr
\+HeD10   & He-det.      &  0.8+0.22  &  .036  &  ---   &  ---  &
                           0.75 (0.1)              \cr
\hline
\+DET2env4 & det.+envelope  &  1.2 + 0.4  &  0.04  &  ---  &  ---  &
                                      0.63                    \cr
\hline
\endtab

Basic quantities of the dynamical models are given in Table 1.
The quantities given
in columns 3 to 7 are:  $M_\star$, white dwarf mass; $\rho_c$, central
density (in CGS) of the white dwarf; $\alpha$, ratio of deflagration
velocity and local sound speed; $\rho_{tr}$, transition density
(in CGS) at which the deflagration is assumed to turn into a detonation;
 $M_{Ni}$, mass of synthesized \ni. The  number in brackets is the amount of
high velocity \ni (Fig. 1).
 For the helium detonations and the envelope (merger) models,
the mass of the C/O core, and the He-layers of the hydrostatic white dwarf
or of the extended CO-envelope, respectively, are given separately.
For all models the initial mass ratio of C/O was assumed to be 1 except
for model DD203c, for which C/O was assumed to be 2/3.

 The models for delayed detonations, pulsating delayed detonations
and envelope (merger) models are selected from the total array of
those models to be the subset with parameters which produce reasonable
multi-color light curves of events in the observed sample of
both normally bright and subluminous SNe~Ia.  Calculations of
spectral evolution or even maximum light spectra do not yet exist
for the whole sample of models, but the selected models are chosen
with composition/velocity profiles that are not in obvious
disagreement with observations (as are the pure detonation models).
For comparison, we note that the structure of the deflagration model
W7 (Nomoto et al. 1984) is similar to that of DD201c.
The helium-detonation models have been advocated in the literature
(Livne \& Arnett 1990, Woosley and Weaver 1994, Arnett 1997)
as an explanation for both normally bright and subluminous
SNe~Ia.  Sub-Chandrasekhar progenitors would, in principle, be easier
to produce from stellar evolution than Chandrasekhar-mass progenitors.
There are problems with both the light curves and spectra of
this class of models. The light curves are generally
too blue  (H\"oflich et al. 1997),
and spectra are dominated by Co/Ni.  In addition, the normally
bright models show excessively weak Si II at maximum light and the
subluminous versions of the model have the Si moving at too low
an expansion velocity (H\"oflich et al. 1997, Nugent et al. 1997).
Despite these objections, it is important to investigate this
class of models and subject it to observational check in as
many ways as possible.  As we will argue below, future $\gamma$-ray
observations can provide a critical check of this class of models.

\section{Spectral evolution}

 The evolution of the hard X-ray and $\gamma $-ray spectrum
of the delayed detonation model DD202c is shown in Figs. 2 and 3
(note the changing scale on the vertical axis).
The spectrum and its evolution is dominated by
nuclear lines and the decay  $^{56}Ni \rightarrow ^{56}Co  \rightarrow
^{56}Fe$, respectively.
During the first 2 weeks, \ni lines dominate. The continuum is formed by
Compton scattering produced by ``down" scattered line photons.
 Below 0.3 MeV (Fig. 3), the spectrum is dominated by the Compton
scattering continuum with the exception of a
very strong \ni line at 0.16 MeV which can be seen for at least
40 days after the explosion. This line
may be observable with a new generation of hard X-ray instruments (see section
5).
 \begfig -.2cm
 \psfig{figure=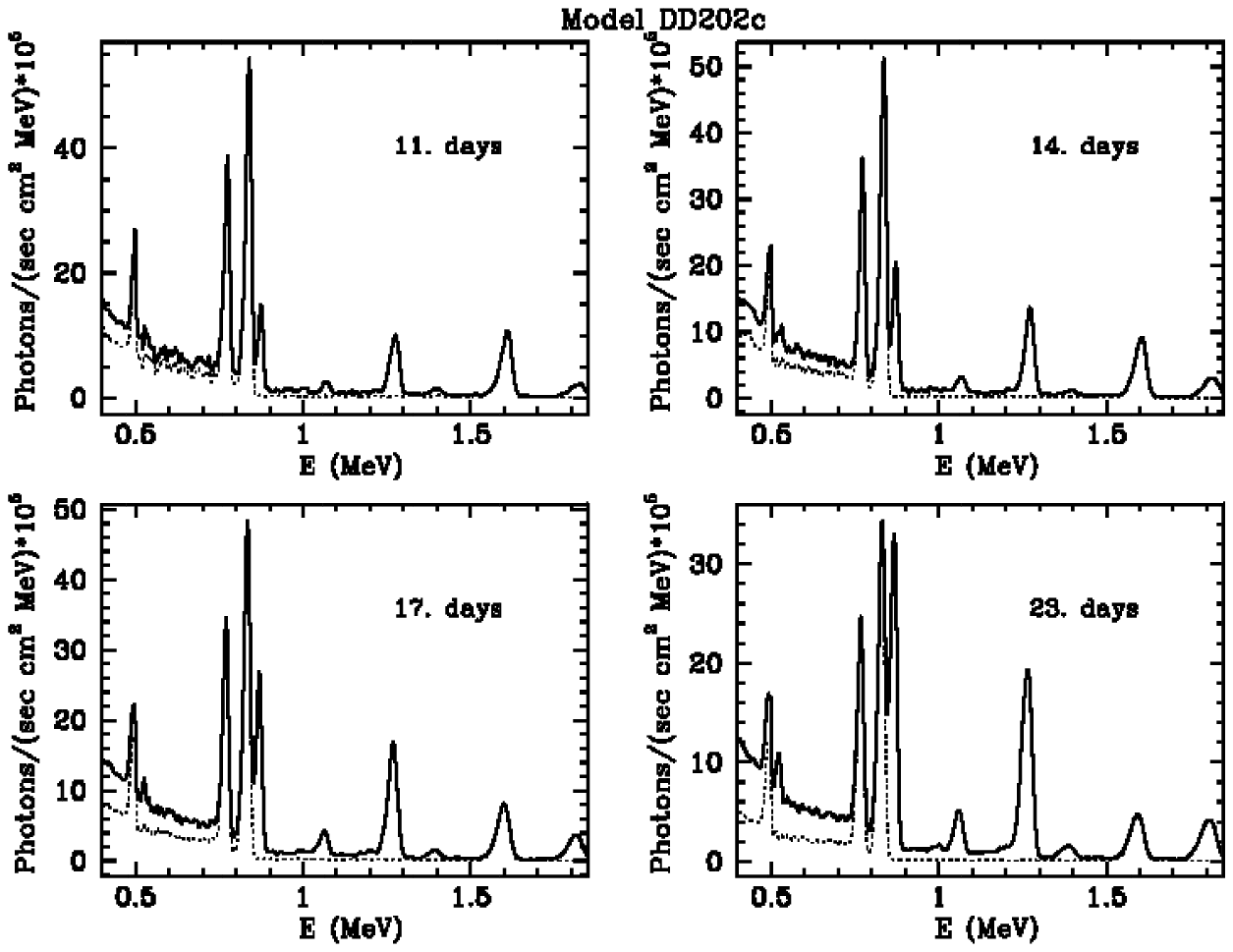,width=13.6cm,rwidth=14.5cm,clip=,angle=270}
 \psfig{figure=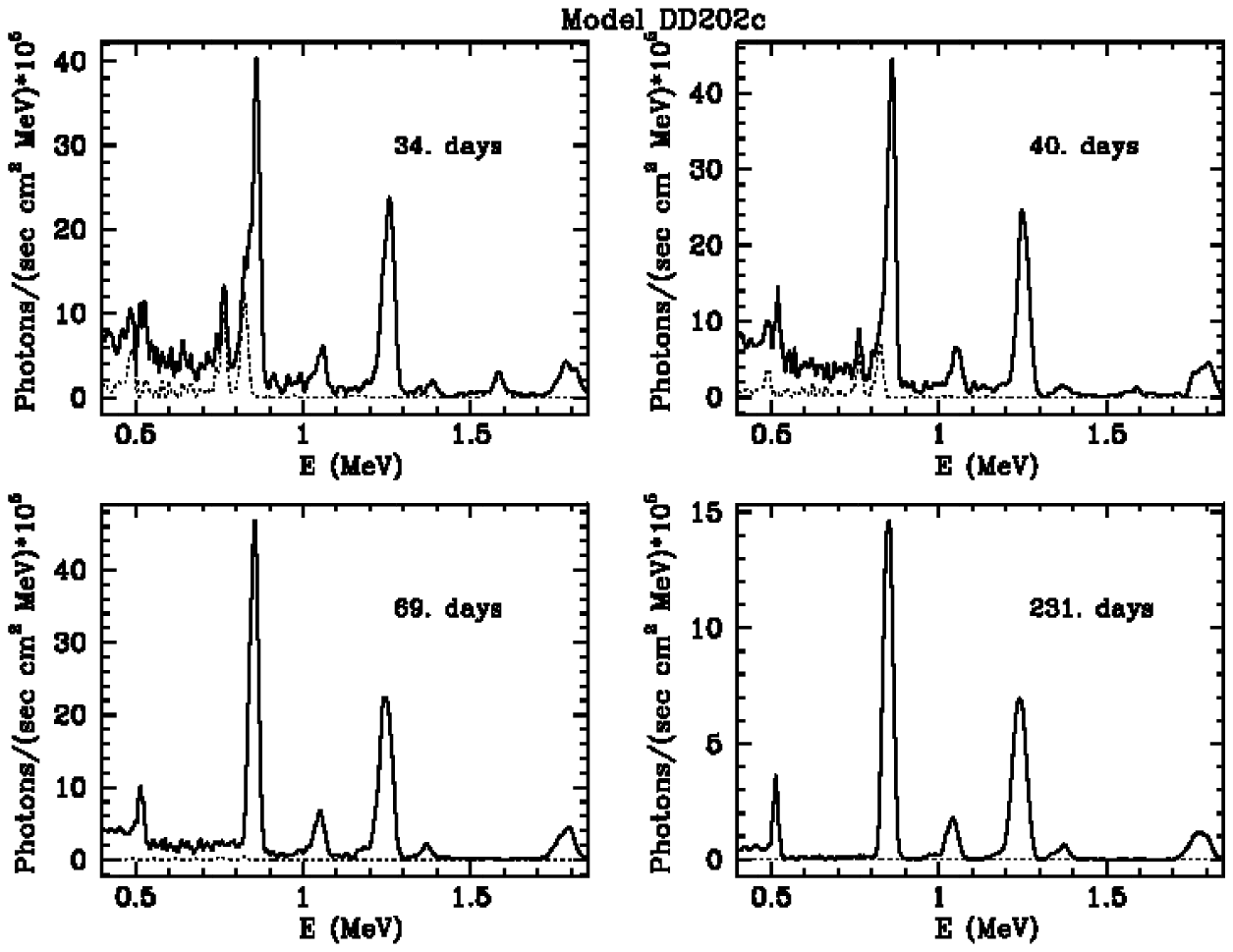,width=13.6cm,rwidth=14.5cm,clip=,angle=270}
\figure{2}{ Evolution of the gamma-ray spectra of the delayed detonation model
DD202c.
The dashed lines corresponds to the contribution of $^{56}Ni$.
The strongest lines are \ni(0.48MeV), the positron decay line at 0.51MeV,
\ni (0.81MeV), \co (0.84MeV) and \co (1.23MeV).
}
\endfig
 \begfig -.2cm
 \psfig{figure=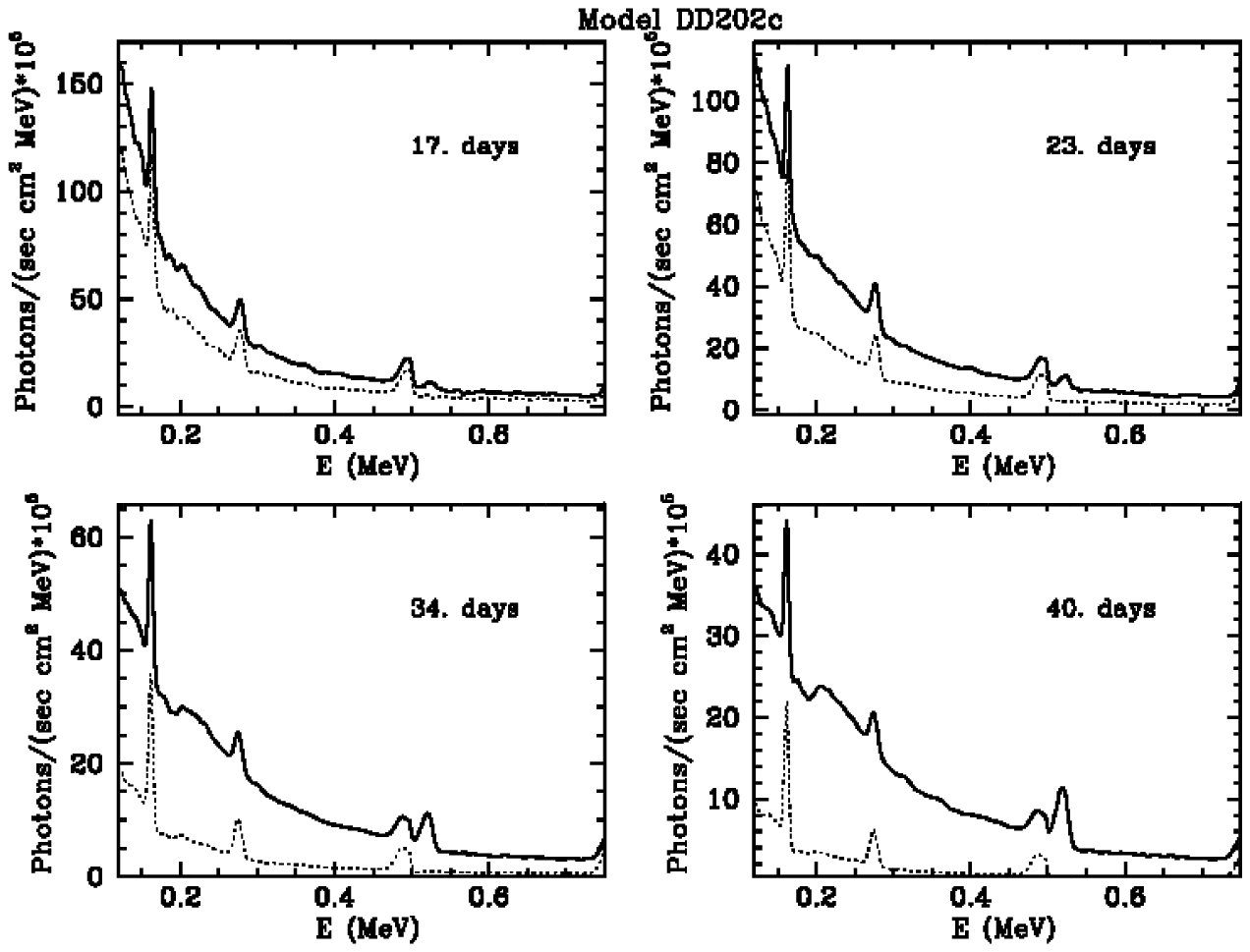,width=13.6cm,rwidth=14.5cm,clip=,angle=270}
\figure{3}{Same as Fig. 2, but for lower energies. The strongest lines are
the \ni lines at 0.158, 0.26 and 0.48 MeV and the positron decay line at 0.51
MeV.
}
\endfig

After two weeks or so, the \co lines become the most prominent line features.
With increasing time, the Compton optical depth decreases and, eventually,
the Compton continuum becomes weak (lower panels in Fig. 2).
Due to the two- and three-photon decay of ortho- and para- positronium,
a strong 511 keV line and a flat continuum below 511 keV is produced.
 After about 6 weeks, the overall spectral distribution remains unchanged.
 \begfig -.2cm
 \psfig{figure=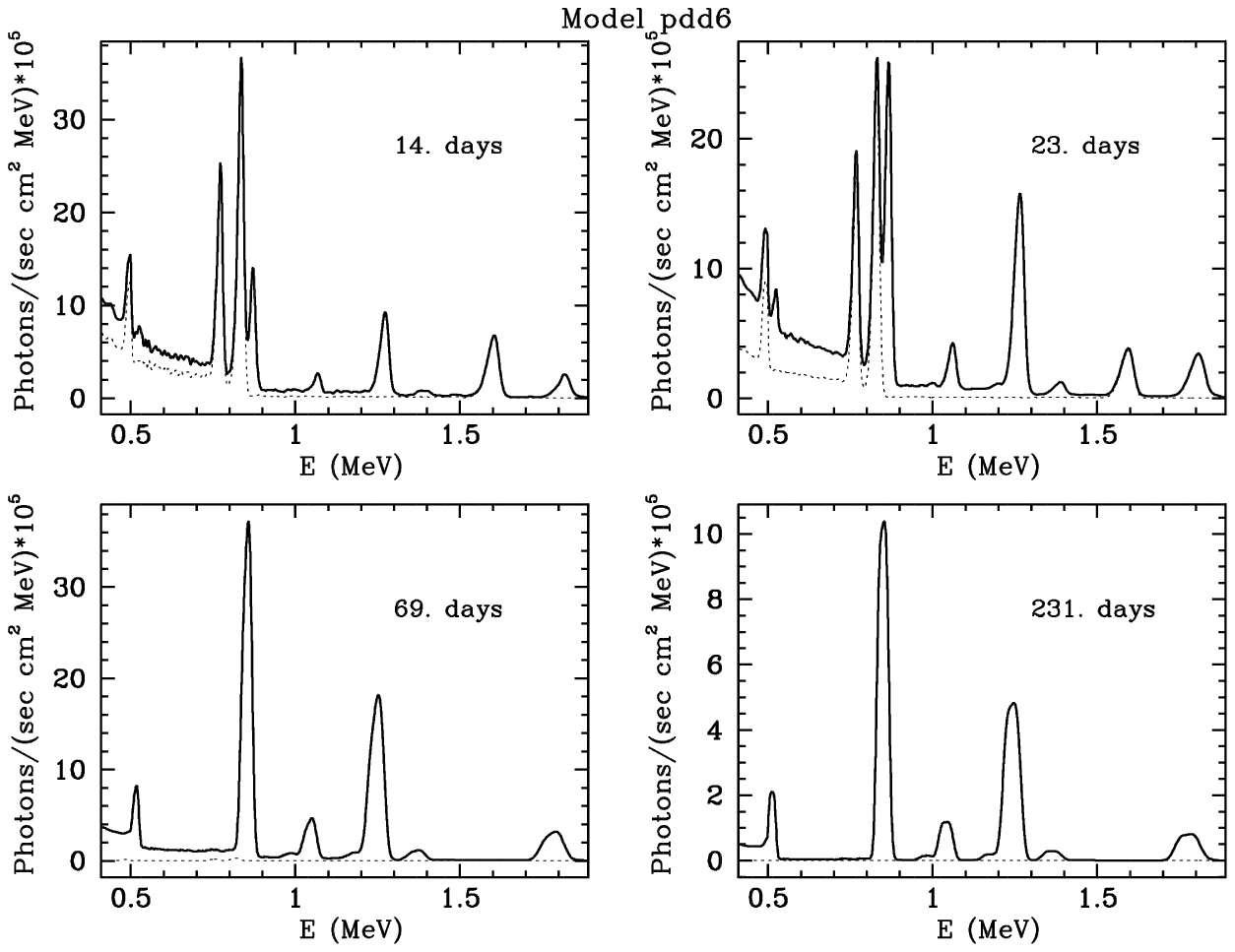,width=13.6cm,rwidth=14.5cm,clip=,angle=270}
 \psfig{figure=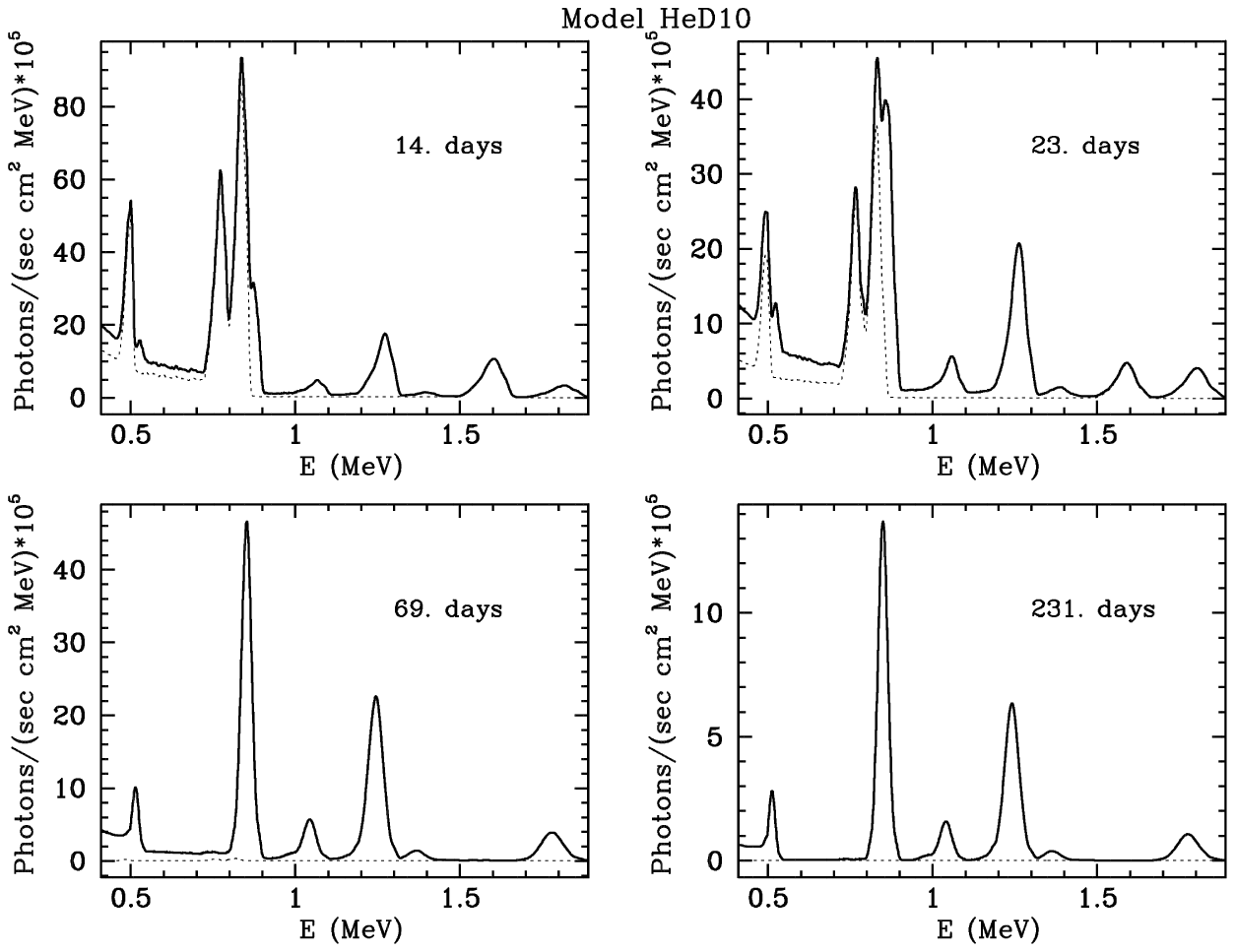,width=13.6cm,rwidth=14.5cm,clip=,angle=270}
\figure{4}{ Same as Fig. 2, but for the
 normal bright supernova models PDD6 and HeD10}
\endfig

The spectral evolution of other evolutionary and dynamical
scenarios is shown in Fig. 4 which illustrates model PDD6,
a pulsating delayed detonation model which is an alternative
for a normally bright SNe~Ia and HeD10, which is an
example of a bright helium shell, double-detonation model.
Qualitatively, the profiles look rather similar for all models
(including deflagrations like W7, H\"oflich et al.
1992), with the noticeable exception of the helium detonations.
For the Chandrasekhar mass or merger models, the model spectra
differ mainly at late times when the ejecta are transparent.
At this phase, the central hole in the \ni distribution
results in flat line profiles.  Note the difference between
model DD202c in Fig. 2 and model PDD6 in Fig. 4 at 231 days,
where the latter shows a distinct flat top to the lines.
These differences give promise for constraints on
the central density of the progenitor, the initial
deflagration speed and, possibly, mixing processes (H\"oflich et al. 1992).
In the absence of gravitation at the center, the flame speed is
expected to be close to the laminar deflagration
speed and then to accelerate as Rayleigh-Taylor instabilities
take over (Khokhlov 1995, Khokhlov et al. 1997ab, Niemeyer \& Hillebrandt
 1995,  Niemeyer \& Woosley 1997).
Late time $\gamma$-ray spectra provide valuable information on the velocity
distribution
of radioactive material in the expanding envelope
 and, thus, they probe the central density of the WD.
To differentiate between different Chandrasekhar mass
and merger models requires high resolution spectra corresponding
to about 1000 km sec$^{-1}$  because
the maximum \ni velocity varies from about 2000 to 3000 km $s^{-1}$
for models which are able to reproduce observations. The
\ni abundance typically changes by about a factor of 80 percent
over a velocity range of 1000 km  $s^{-1}$. Thus, signal to noise ratios of
the order of 5 are required to separate the various effects.
An alternative, less observationally demanding method is discussed below.

 Helium detonations show significantly  different $\gamma$-ray spectra
and spectral evolution compared to the Chandrasekhar mass and
merger models.
 At all phases, the helium-detonation models can thus be
distinguished from both Chandrasekhar mass and merger models
with appropriately sensitive observations.
The \ni at high velocities from the outer detonated
helium shell results in very broad profiles beginning a few days
after the explosion. Two to three weeks after explosion,
near or after maximum optical light, the escape probability of
$\gamma$-rays from the central, low velocity \ni
region is sufficient to produce a narrow component atop the broad
component (see panel 2 for model HeD10 at 23 days in Fig. 4).
Depending on the total amount of \ni, the central component
dominates after about one month and, eventually, a narrow
strongly peaked line remains with low amplitude, but very
extended wings.
 At late times, the amplitude of the high velocity component is lower
than the low velocity component by almost an order of
magnitude even though the \ni mass is lower by a factor of only about 7
in the normally bright model (HeD10) and 3 in the subluminous model
(HeD6).  The reason is that the the outer \ni is smeared out over about twice
the velocity range, thus diluting the flux at a given energy.

 \begfig -.2cm
 \psfig{figure=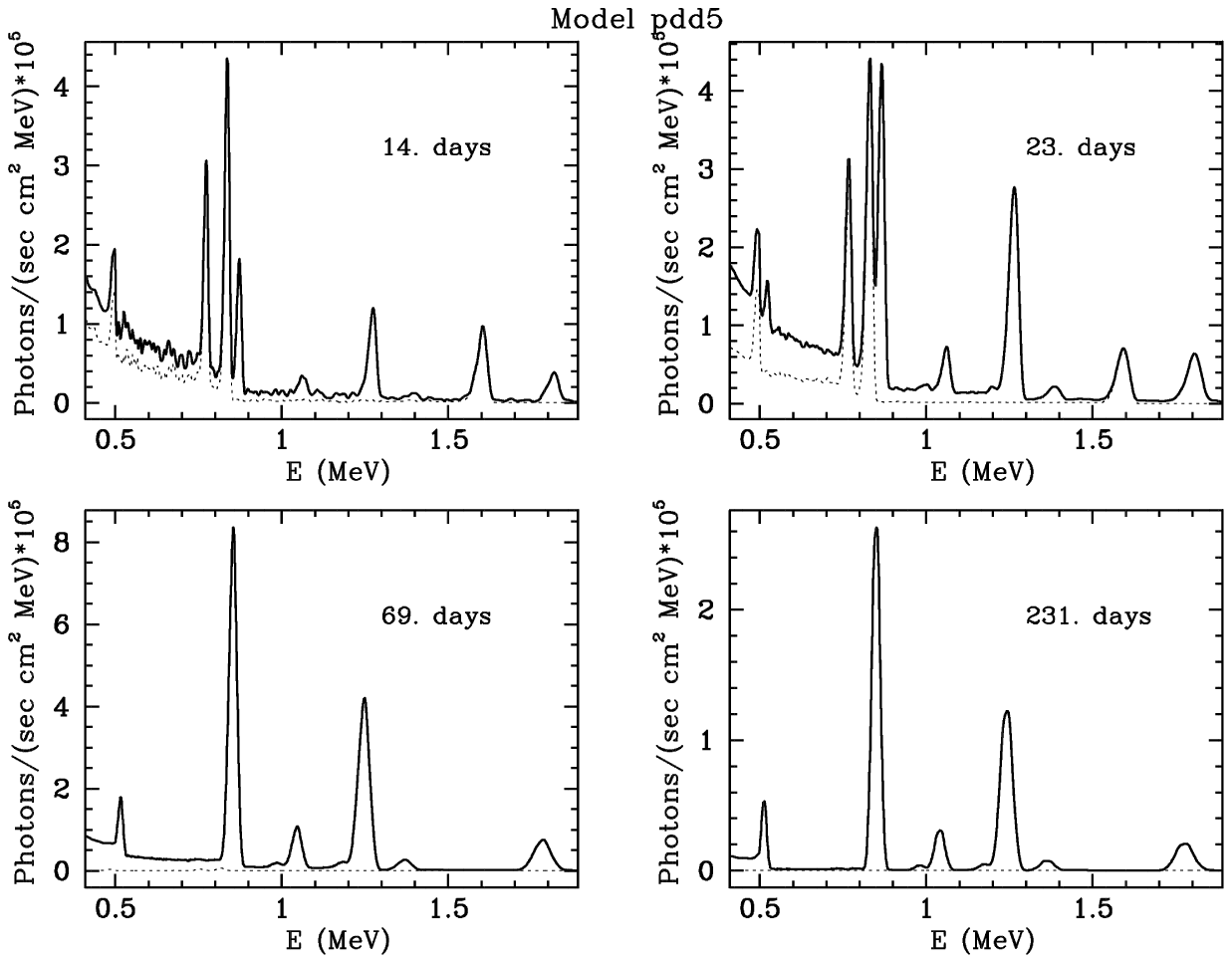,width=13.6cm,rwidth=14.5cm,clip=,angle=270}
 \psfig{figure=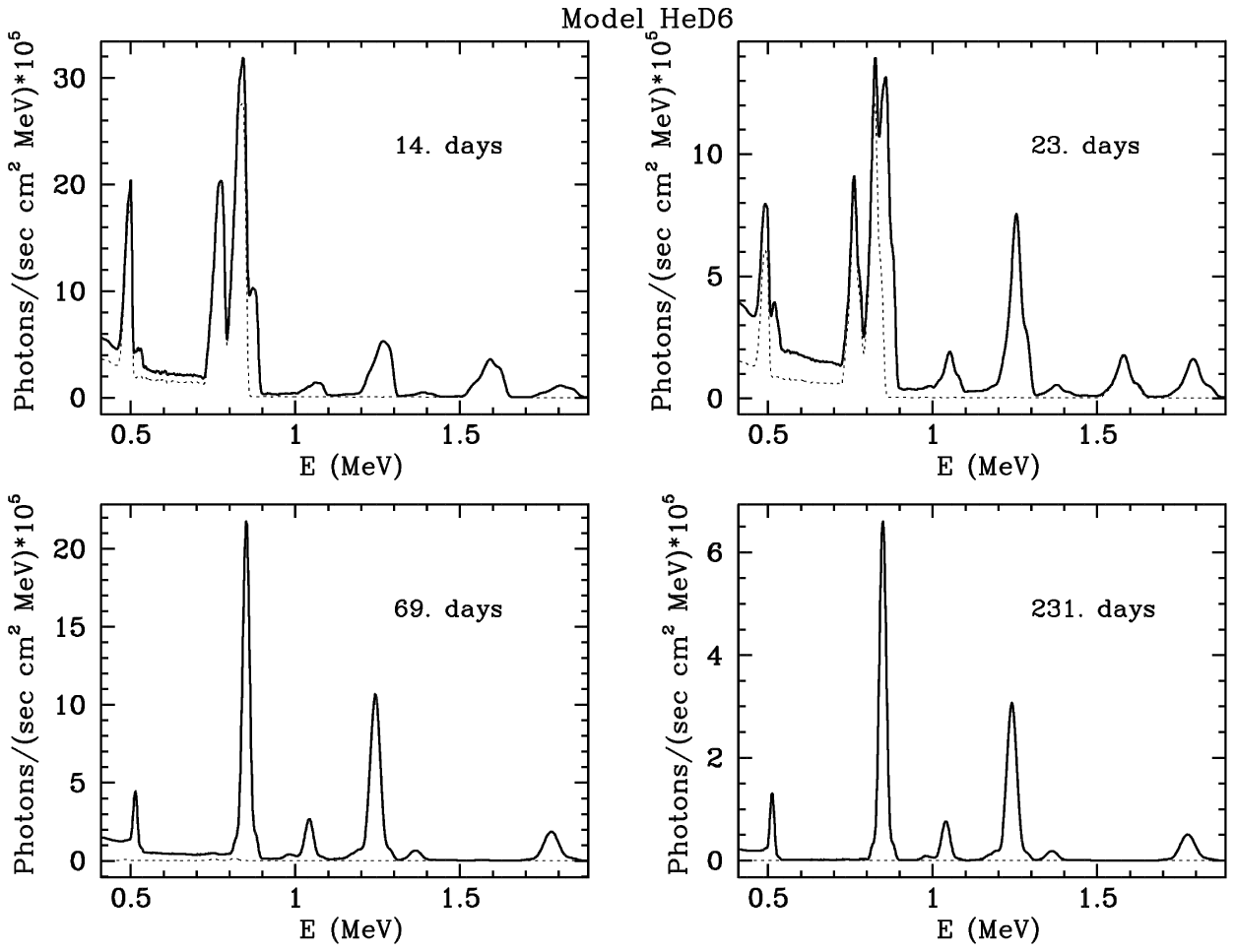,width=13.6cm,rwidth=14.5cm,clip=,angle=270}
\figure{5}{ Same as Fig. 2, but for the subluminous supernovae models PDD5 and
HeD6.}
\endfig

 The differences between the Chandrasekhar mass and sub-Chandrasekhar
mass models are even more defined when the subluminous versions
of each are compared.  In massive white dwarf models, the mean expansion
velocity of the \ni decreases with the decrease in total \ni mass.
This means that the subluminous models that have less nickel
according to the dynamical models (H\"oflich, Khokhlov,
and Wheeler 1995) have narrower lines
compared to the models for the normally bright SNe~Ia (see also Fig. 7).
For sub-luminous helium detonations at early epochs, the \ni at
high velocities makes the dominant contribution to the flux
in a given \ni line.  Consequently, the \ni lines are very broad
and nearby lines are blended, unlike the Chandrasekhar-mass models.
The mean velocity of the inner \ni in the brighter helium-detonation model
(HeD10 in Fig. 4) is higher than that for the less luminous model
(HeD6 in Fig. 5).  The velocity profiles of the outer helium
layer are similar for the two models.
As a result, at late times, the broad ``shoulder" remains
more visible in the $\gamma$-ray spectra of the subluminous helium
detonation models, as can be seen by comparing the spectra at
231 days of model HeD10 in Fig. 4 with model HeD6 in Fig. 5.
This ``shoulder" in the subluminous models is in especially
strong contrast to the late-time line profile of the subluminous
Chandrasekhar-mass model as shown in Fig. 5.

In principle, the outer He-\ni layers of
helium-detonation models may also be observable in the
optical and IR for the first few months, because non-thermal
electrons                            can provide excitation
of the corresponding lines (H\"oflich et al. 1997).
In the absence of a magnetic field, the positrons can freely escape
after a few months and the optical and IR emission would
decline.   With a tangled magnetic field and
a continued trapping of the positrons, this associated1
optical and IR emission could continue more strongly to later epochs.
Early optical and IR spectra may thus provide an additional
means to detect the He-\ni layer, but this possibility is
currently controversial.  Whereas two groups (H\"oflich et al. 1997,
Nugent et al. 1997) find that helium detonations give maximum light
spectra and light curves that do not reproduce the observations of SNe~Ia,
Pinto \&  Eastman (1995, private communication and conference talk in Spain)
argue that agreement is possible.  Consequently, $\gamma $-rays
with their straightforward interpretation must be regarded as
the best way to ultimately clarify the nature of SNe~Ia.
 \begfig -.2cm
 \psfig{figure=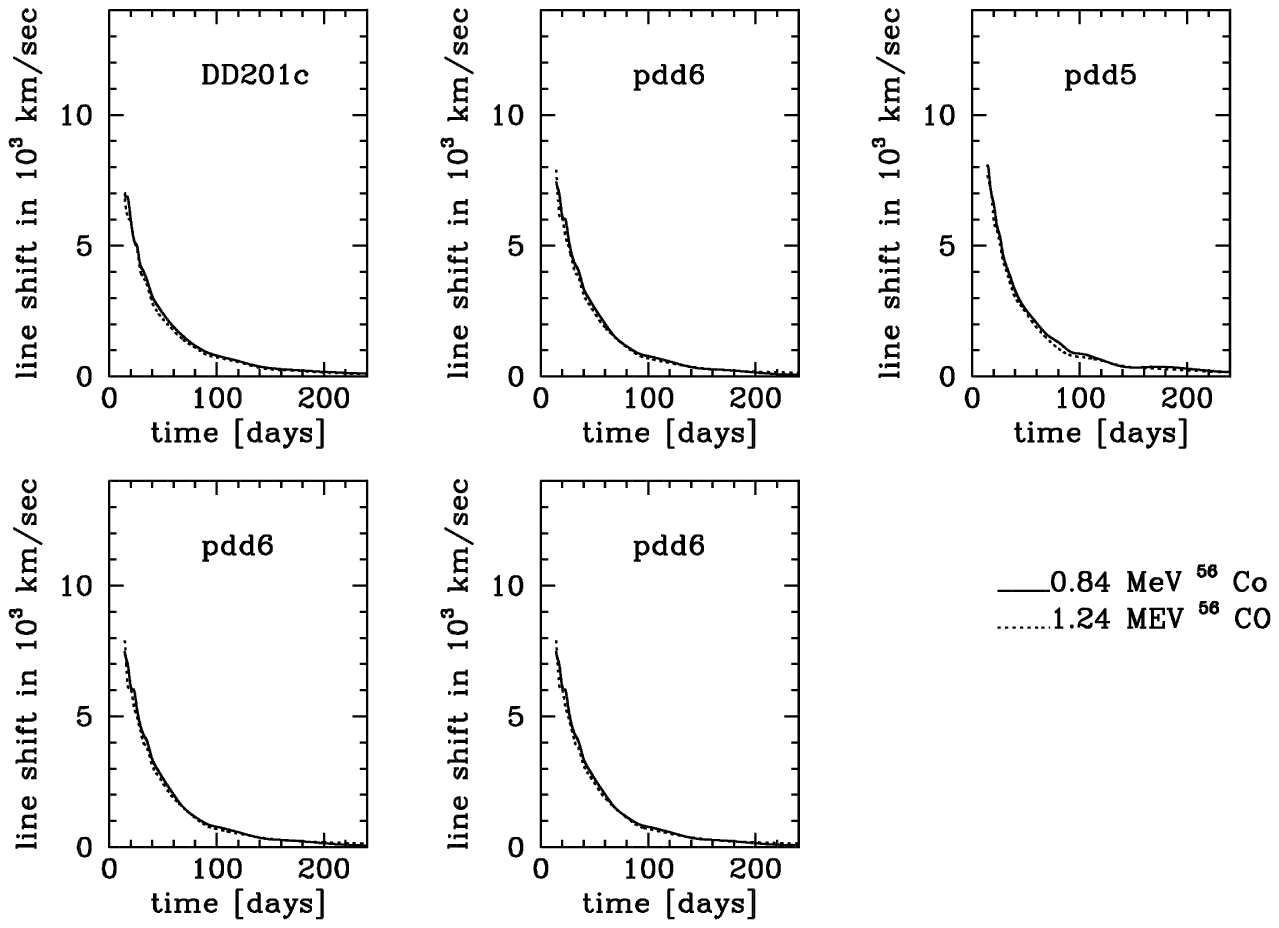,width=13.6cm,rwidth=14.5cm,clip=,angle=270}
 \psfig{figure=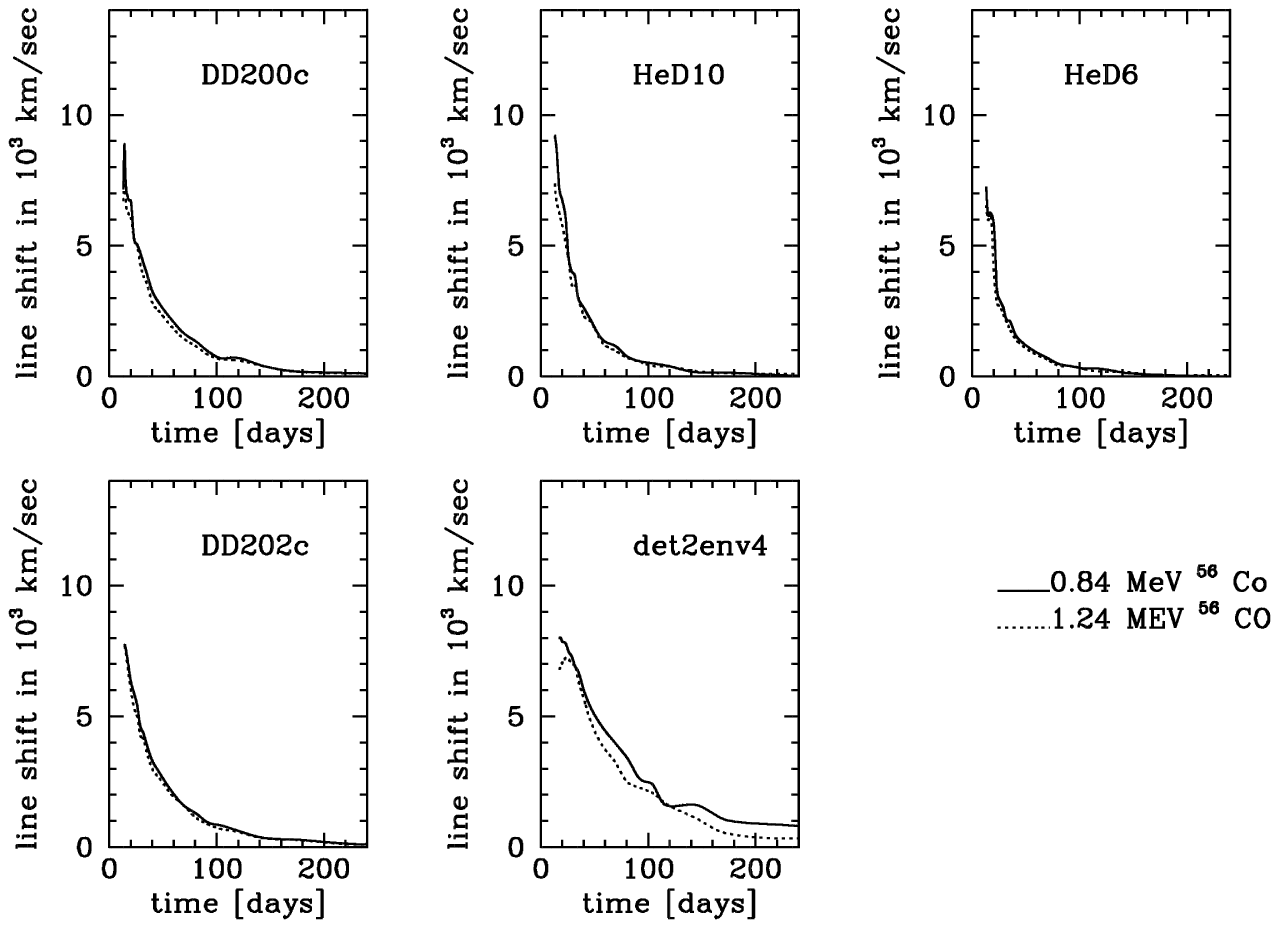,width=13.6cm,rwidth=14.5cm,clip=,angle=270}
\figure{6}{
Time evolution of the line shift relatively to the rest frame.
Note that the line shift is insensitive to the specific line considered.
}
\endfig

 \begfig -.2cm
 \psfig{figure=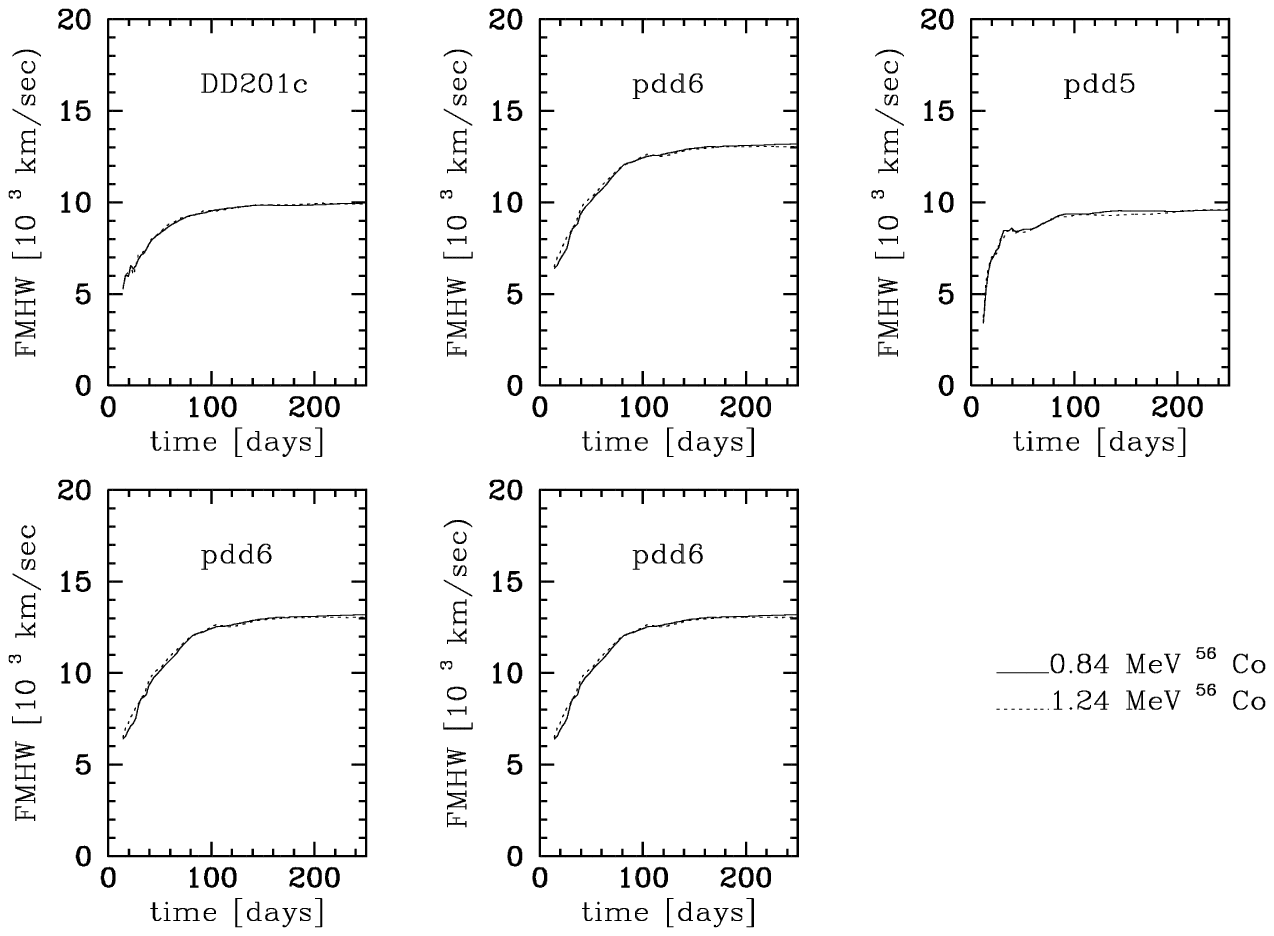,width=13.6cm,rwidth=14.5cm,clip=,angle=270}
 \psfig{figure=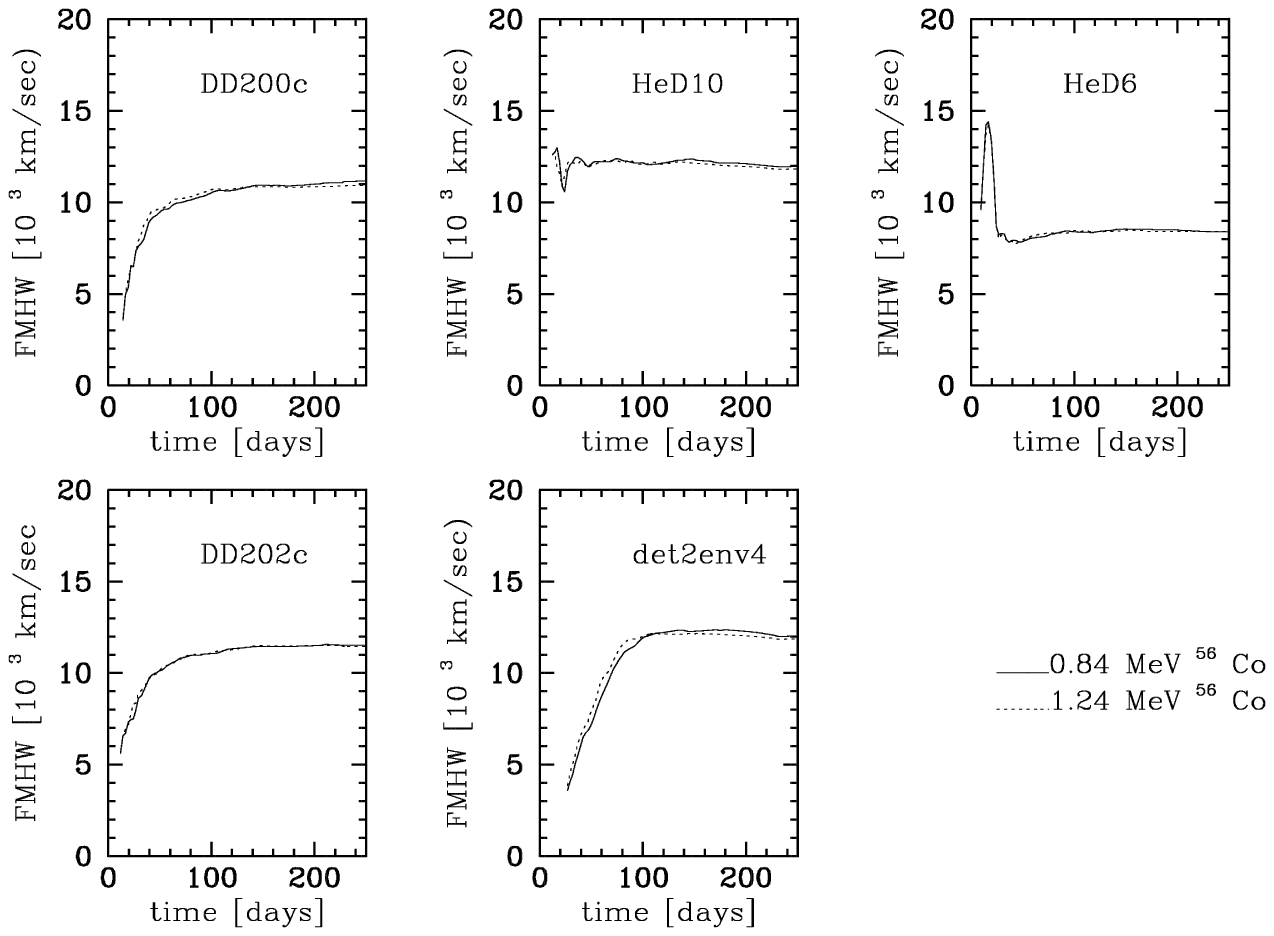,width=13.6cm,rwidth=14.5cm,clip=,angle=270}
\figure{7}{
Time evolution of the Full Width Half Maximum (FWHM)
of the 0.84 and 1.24 MeV $^{56}Co$ lines. Note that the  FWHM is
insensitive to the specific line considered.}
\endfig

\subsection {Quantitative Analysis of Line Profiles}

In order to provide specific guidance for observational missions,
it is appropriate to present more specific detail of the
properties of the $\gamma$-ray line profiles given in the previous
section.  Of special importance are the line widths and the
the displacement of the line with respect to the
rest wavelength due to the optical depth and structural effects
in the models.  The time evolution of the displacement of the
line for a variety of models is given in Fig. 6 for the two
strong lines of \co at 0.84 and 1.24 MeV.  Fig. 7 shows the
evolution of the width (Full Width Half Maximum) of the same \co lines.
Because the Compton cross section decreases slowly with energy
and expansion effects dominate, the evolution of the line displacement
and width is relatively insensitive to the specific line.
Figs. 6 and 7 show that the rate of decline of the blue shift of
the 0.84 MeV \co line is only slightly faster than that of the
the 1.24 MeV line and that the evolution of the line widths
are virtually identical for all models.
This common behavior might make it easier to derive constraints
from low signal to noise data from missions like CGRO and INTEGRAL.
The line shift is critical for a proper analysis of
proposed high sensitivity but narrow wavelength band instruments
such as Laue telescopes.

During the first few weeks, the maxima of all $\gamma$-ray lines
are blue shifted by a velocity that is typically 5000 to 8000 km sec$^{-1}$
with respect to the rest wavelength because
the ejecta are optically thick to $\gamma$-rays during this
epoch and the flux arises from the approaching side.
Photons escape essentially radially from the
\ni region that is expanding towards the observer.
The blue-shift is comparable to the intrinsic line width,
another measure of the kinematic velocity of the radioactive
matter.  The rate of decrease of the blue-shift
depends on the optical depth of the \ni region and of the
matter above it.  Consequently, the displacement tends to drop most
rapidly with time in the sub-Chandrasekhar models HeD6 and HeD10,
and most slowly in the envelope models such as DET2env4 or
the pulsating delayed detonations such as PDD6 (cf Fig. 6).
The evolution of the blue shift during the early epochs,
prior to 30 days after the explosion, roughly two weeks after optical maximum,
might allow discrimination of Chandrasekhar and sub-Chandrasekhar mass
models of the subluminous variety of SNe~Ia as can be seen
by comparing the predicted displacement evolution for
models PDD5 and HeD6 in Fig. 6.
With the exception of early time observations,
the line shift is not a powerful discriminant between
helium detonation and delayed detonation models.
Careful observations of the blue-shift over the first
150 days may, however, provide such a test for models with
masses larger than $M_{Ch}$. If observed, a slow
evolution of the blue shift could prove
the existence of  the shell-like structures expected for mergers.

Figure 7 shows that $\gamma$-ray line widths can provide detailed
discrimination of models for instruments with spectral resolution
of order of $\lambda / \delta \lambda \sim  200  - 300$.
Since the line width depends on the velocity distribution
of the radioactive material but is insensitive to the wavelength,
the results for the 0.84 and 1.24 MeV lines of \co
given in Fig. 7 are quantitatively valid for other lines as well.
As shown in Fig. 7, the line width for most models
increases on time scales of 10 to 100 days
because radioactive material from an increasing velocity range
becomes visible as the $\gamma$-ray optical depth declines.
Since more \ni generally implies less shielding by overlying layers
and, thus, higher escape probability at early times within each series of
models, the rate of increase of line width is smallest for
models with lower \ni masses  (e.g. DD201c vs. DD202c).
For all models, the rate of increase in line width
is also smallest for models with ``well hidden" \ni.
This is the case if the model naturally produces a definite
shell-like structure such as PDD6, or if the total mass is larger
than $M_{Ch}$ as for the merger models like DET2env4.

The behavior of the sub-Chandrasekhar mass models with detonating
outer helium shells is also conspicuously different in terms of the
line width behavior.  The lines are very broad beginning
soon after the explosion, because of the high expansion velocity
of the outer \ni produced by the helium detonation.
The line width decreases rapidly at about 2 weeks
after the explosion.  The rapid decline in line width in the
helium-detonation models can be understood in terms of the
line profiles illustrated in Figs. 4 and 5 as being due to the
increased emission from the narrow inner \ni component (see above).
With increasing time, this narrow component from the
centrally-produced \ni dominates. The secondary minimum occurs approximately
when the monochromatic flux due to the central \ni at a
low projected velocity dominates the flux of the high velocity
component shifted by one Doppler width as
the contribution of the high velocity component becomes
insignificant due to the larger velocity spread and  smaller mass.
For  normally bright helium-detonation models such as
HeD10, the central \ni mass exceeds that of the outer
region by  more than a factor of $\approx 7$ whereas
for subluminous models such as HeD6, this factor is less
than 3 (Table 1).
Consequently, for subluminous helium detonations, the early
spike in the line width is especially pronounced
and should be a clear diagnostic for such models.

After about 50 to 150 days, depending on the model,
the line widths approach asymptotic values
as the envelopes become transparent to $\gamma$-rays.
In general, the asymptotic value of the FWHM line width corresponds
neither to the expansion velocity of the matter within which
\ni dominates the abundances nor to the mean velocity of the \ni .
Rather, it measures the projected expansion velocity
at which half the amount of radioactive material is located
relative to the material seen with zero expansion velocity.
For models with centrally concentrated \ni, the FWHM depends mainly
on the amount of \ni and the size of the central hole
of the \ni distribution. The former can be independently determined
by the line flux at late times (see next section). Therefore, the
line width provides a unique opportunity to probe the
very central region and to get an insight into the
initial conditions at the time of the explosion (see above), namely, the
density where the burning starts and the central density of the white dwarf.

\begfig -.2cm
 \psfig{figure=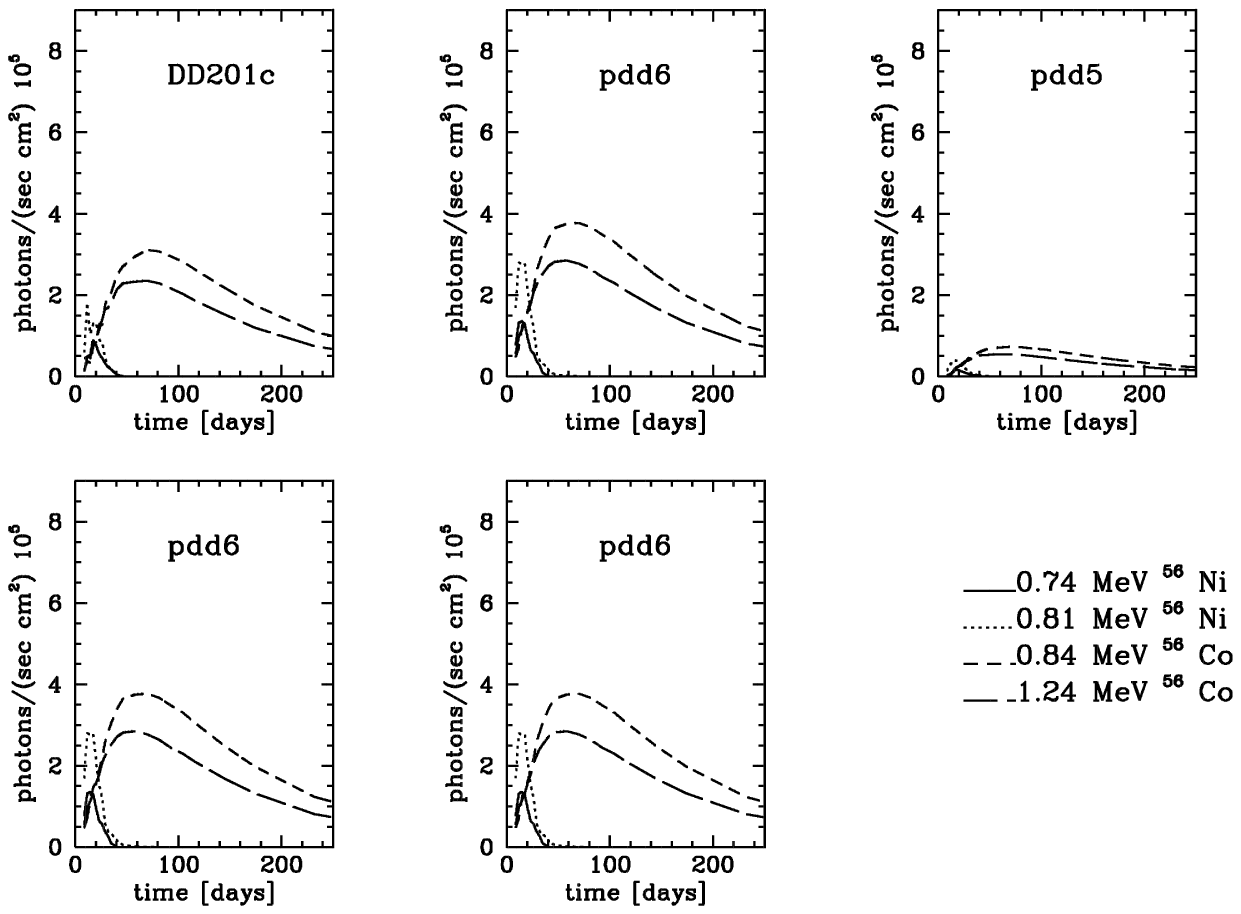,width=13.6cm,rwidth=14.5cm,clip=,angle=270}
 \psfig{figure=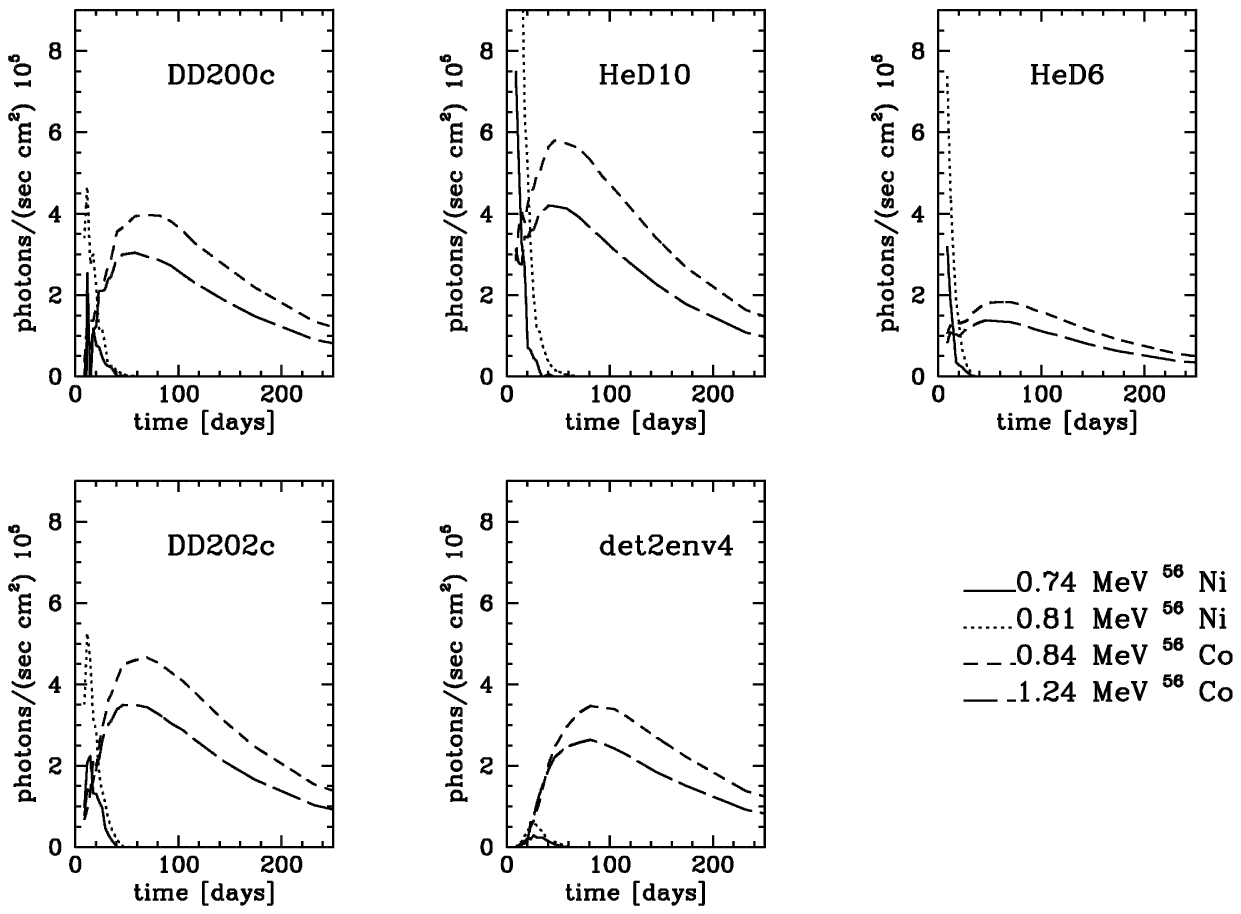,width=13.6cm,rwidth=14.5cm,clip=,angle=270}
\figure{8}{Time evolution of the
 integrated line fluxes of the most prominent gamma-ray lines of $^{56}Ni$ and
$^{56}Co$ as predicted by theoretical models (see Table 1) assuming a distance
of
10Mpc}
\endfig

\begfig -.2cm
 \psfig{figure=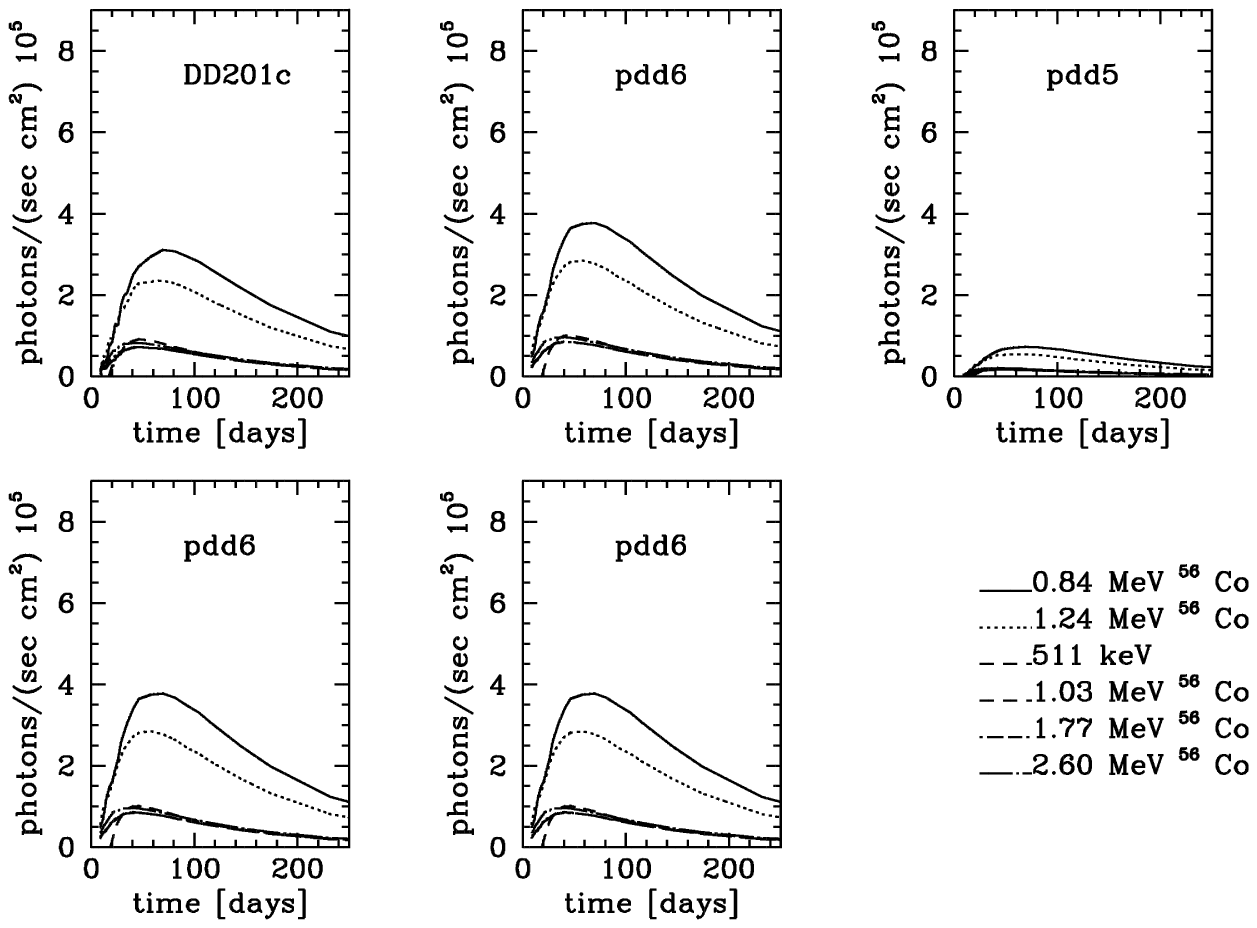,width=13.6cm,rwidth=14.5cm,clip=,angle=270}
 \psfig{figure=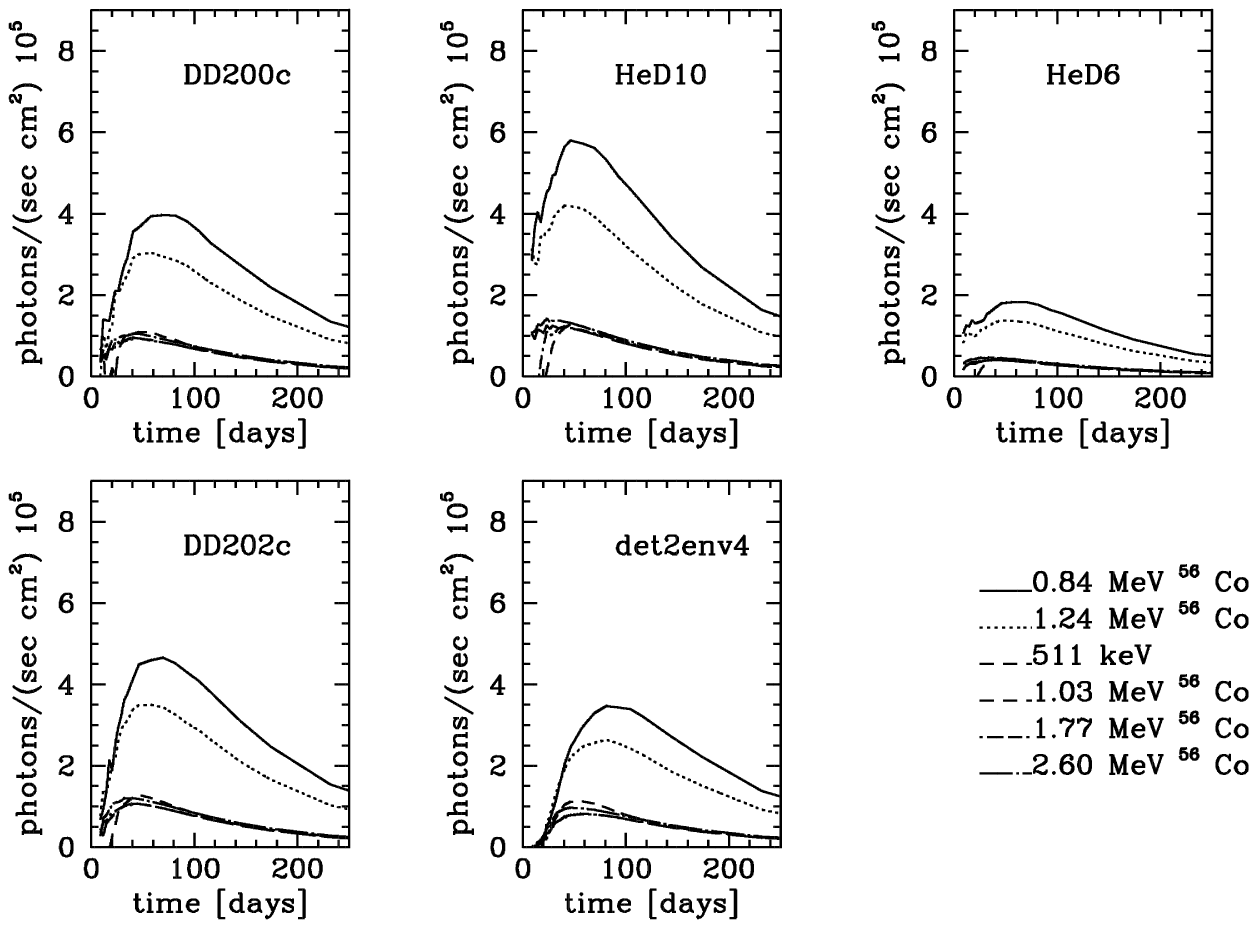,width=13.6cm,rwidth=14.5cm,clip=,angle=270}
\figure{9}{Same as Fig. 8, for $^{56}Co$, but including weaker lines.}
\endfig

 \begfig -.2cm
 \psfig{figure=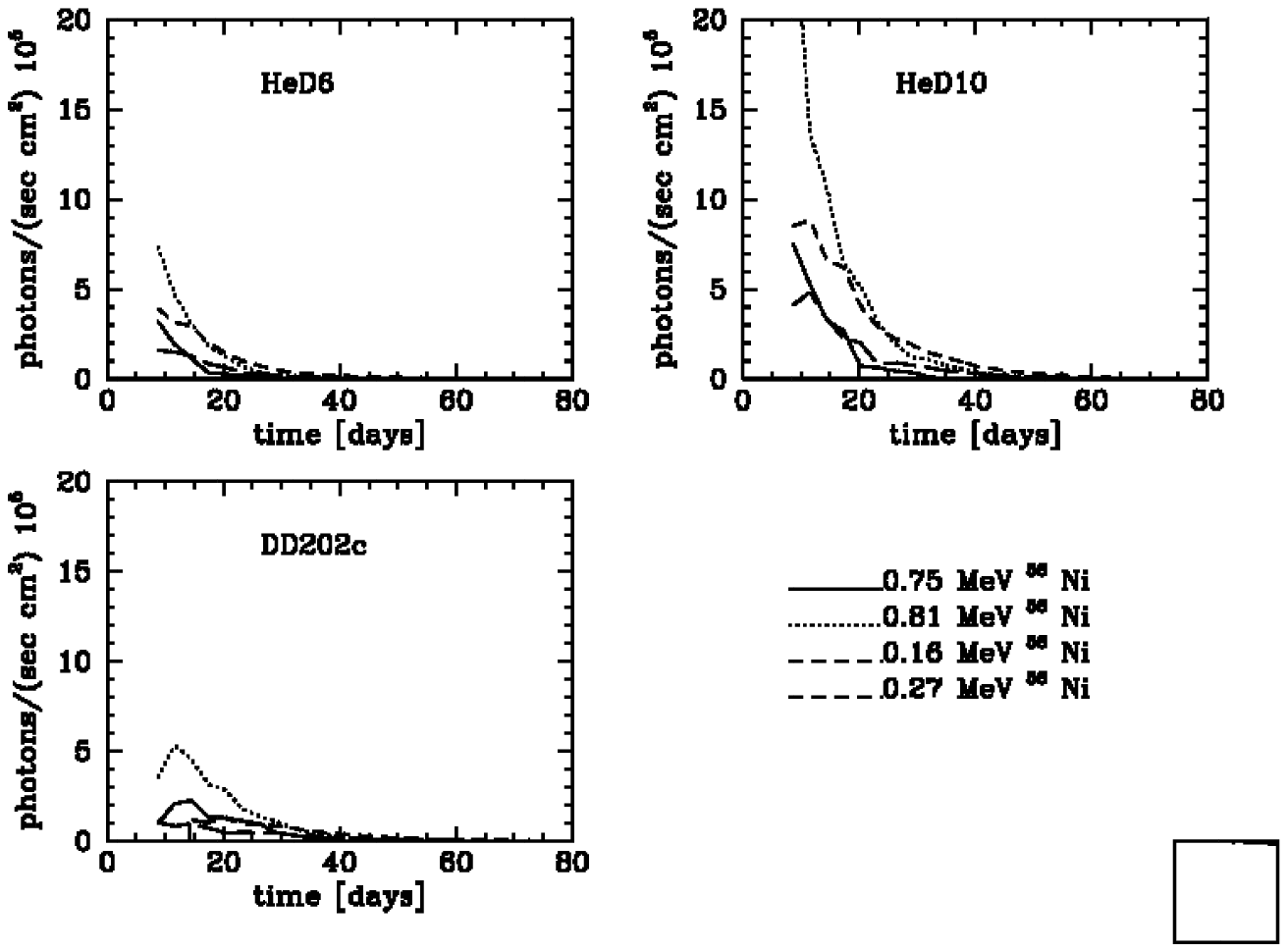,width=13.6cm,rwidth=14.5cm,clip=,angle=270}
\figure{10}{Same as Fig. 8, for $^{56}Ni$.}
\endfig

\section {Gamma-ray Line Fluxes and Light Curves}

The $\gamma$-ray line fluxes and resulting light curves provide
another diagnostic to choose among models for the explosion.
The information available will depend on whether one knows,
from other means, the distance to the source.
Fig. 8 presents the line flux light curves for the two strongest
lines each of \ni and \co for a range of relevant models.
Fig. 9 gives the same information for \co, including some of
the weaker lines and Fig. 10 does the same for \ni.

We first consider the information contained in distance-independent
quantities, such as flux ratios.
Fig. 8 shows that the helium-detonation models with radioactive
material close to the surface are predicted to have
an especially strong line flux at early times.
For helium-detonation models, the maximum absolute
flux due to the \ni lines exceeds that of the strongest \co lines by about
a factor of 5.  For all other models, the \ni flux remains comparable
to or smaller than the flux in \co lines. This provides a crucial,
distance-independent test for the nature of SNe~Ia.

Line flux ratios for lines of the same radioactive isotope provide
a direct measure of the Compton scattering beyond the region
containing the radioactive species and thus another way to distinguish
different scenarios and to separate explosions of sub-Chandrasekhar and
Chandrasekhar mass progenitors.  This is illustrated in
Figs. 9 and 10.  Depending on energy and model, the fluxes of
\ni lines peak between 10 and 20 days after explosion and those
from \co peak between 50 and 90 days.

Figures 3 and 9 show that near and after optical maximum,
the \ni lines at energies below 500 keV have strengths comparable
to the strongest $\gamma$-ray lines at .75 and .81 MeV.
The positron annihilation line at .511 MeV is  only
moderately weaker than other lines in the MeV range.
These results raise the possibility, depending on the sensitivity
curve of the instrument, that low energy \ni and the 0.511 MeV lines
may be detectable more easily than lines in the MeV range.
At early times, all positrons are trapped (Colgate et al. 1997)
and so the results do not depend on model-dependent assumptions
that would alter the results of positron trapping at later times.
\begfig -.2cm
 \psfig{figure=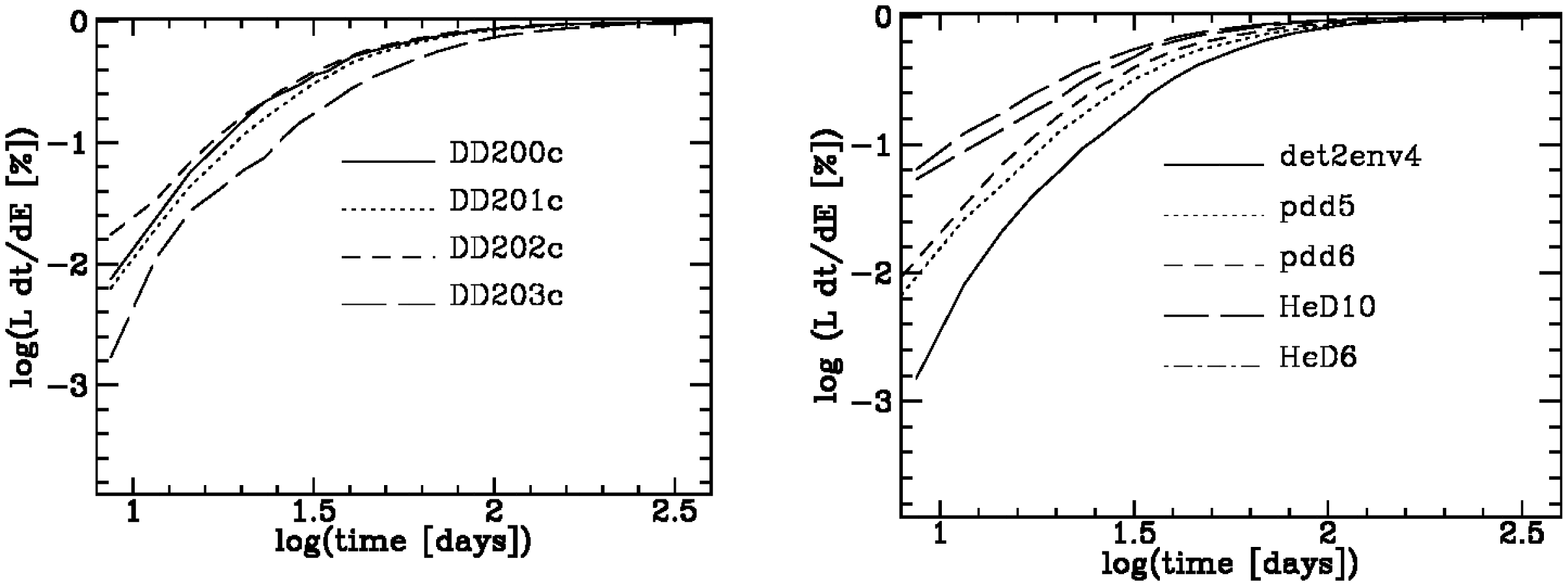,width=13.6cm,rwidth=14.5cm,clip=,angle=270}
\figure{11}{Luminosity above 10keV relative to the energy production by
radioactive
decays.
}
\endfig

Measurement of line fluxes at both early and late times provides
another distance-independent means    to test models.
Fig. 11 presents the ratio R between the total $\gamma$-ray luminosity
and the total instantaneous energy release by radioacive decay.
 The ratio R measures the transparency of the expanding
matter above the region of radioactive elements.
It is very model dependent at early times, but
eventually approaches a value of unity 4 to 6 months after explosion.
Models with high masses and shell-like structures show small R
at early times whereas the helium-detonation models with masses less
than $M_{ch}$ show the highest line  ratios.

 \begfig -.2cm
 \psfig{figure=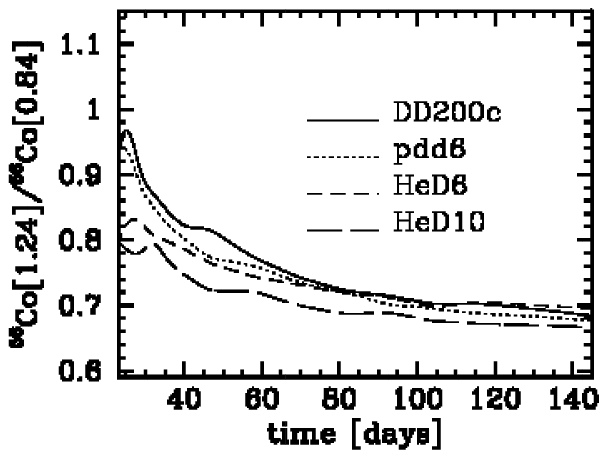,width=11.6cm,rwidth=7.5cm,clip=,angle=270}
\figure{12}
{The ratio  between the $^{56}Co[1.24]$/$^{56}Co[0.84]$ lines. This ratio
depends sensitively on the model.}
\endfig

Besides a comparison of the line strengths of \co and \ni,
the change of the line ratios provides important information on
the structure of the envelope.  This constraint comes by
comparing lines at different energies as illustrated in Fig. 12
which presents results for the 1.24 and .84 MeV lines of \co
for a range of models.  By comparing critical line ratios with their
asymptotic values, the Compton depth can be constrained.
Fig. 12 shows that the \co line ratio is sensitive to both time and
the model.  The line ratios depend mainly on the amount of material above
the radioactive layer; however, self-shielding within the radioactive
region is also important.  At early epochs, even the helium-detonation
models with their exposed outer layers of \ni show line ratios
that substantially exceed the asymptotic value (Fig. 12).
Observations during the first 2 to 3 months are most suitable
in this regard. To distinguish helium-detonation models from the
merger scenario and the explosion of massive white dwarfs,
the line ratios must be known to better than $\approx$ 20 percent.
Note that late time observations are not actually required
because the asymptotic value is fixed by the branching ratio (see Table 1).

Other constraints on the models are possible in
the case where the distance is known from, e.g.,
$\delta $ Cepheids  (Saha et al. 1996, Freedman et al. 1996)
or from optical light curves and spectra of (e.g. Hamuy et al. 1996,
HK96, Riess, Press \& Kirshner 1996).
In this case, models can be distinguished by the time evolution
of absolute line fluxes  as illustrated in Figs. 8, 9, and 10.
For the helium-detonation models, the flux in the strongest \ni lines
should be in excess of $10^5$ photons~sec$^{-1}$~cm$^{-2}$ if observed at
a distance of 10 Mpc as the flux peaks between 10 and 20 days
after the explosion.
Late time measurements would allow the determination of the absolute amount
of radioactive material for given distance.

\subsection{ Determination of the Time of the Explosion}

The early, rapid change of the spectra may allow an accurate
measure of the time of the explosion (Clayton et al., 1969).
Photons at similar energies encounter the same
degradation due to Compton-scattering and, thus, the time-dependent
flux ratio of the lines of similar energy does not depend on the model
as shown in Fig. 13.  Two spectral regions are especially suitable
for determining the time of explosion: that at about 0.8
containing the 0.81 MeV line of \ni and the \co
line at 0.84MeV, and that near 0.5 MeV containing the positron
anhilation line at 0. 511 MeV and the \ni line at 0.48 MeV (Figs. 2 and 3).
Fig. 13 shows the ratio of the 0.81 MeV line of \ni in comparison
to the 0.84 MeV line of $^{56}$Co. The ratio of the 0.48 MeV line to
that at 0.511 Mev follows an essentially identical trajectory,
but with an amplitude decreased by a factor of 3.5.
The ratio of the \ni to \co lines at 0.8 MeV has
the advantage of involving strong lines with a large flux ratio,
but its use is restricted to times between  $\approx $
11 and 25 days without being obscured by contributions of Compton
scattered photons of the \co line.
The ratio of the \ni line to the annihilation line at
0.5 Mev is based on lines with more moderate fluxes but, in principle,
it can be used between day 20 and 50 for nearby SNe~Ia.

 \begfig -.2cm
 \psfig{figure=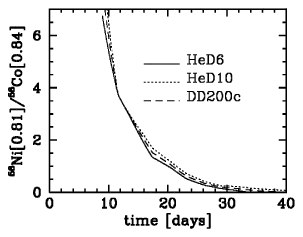,width=11.6cm,rwidth=7.5cm,clip=,angle=270}
\figure{13}
{The ratio  between the $^{56}Ni[0.81]$/$^{56}Co[0.84]$ lines. This ratio
depends sensitively on the time after the explosion, but it hardly depends
on the model. The line ratio between the 0.48 MeV \ni and the 511keV positron
annhilation lines is 3.5.
}
\endfig

\subsection{ Effect of Integration Times}


 Integration times for $\gamma$-ray lines may be on the order of
days to weeks. Therefore it is useful to characterize the
influence of the integration time on the resulting spectrum which
depends on both the mean time and the underlying model.
The model dependence enters because the time evolution of the
absolute fluxes determines the weighting of the line fluxes
over the observed time interval.

 Fig. 14 illustrates the change of the spectrum with integration time
for the specific example of the delayed detonation model DD201c
at about 14 and 23 days after the explosion.
For $\Delta t$ smaller than 10 days, the resulting error in the
line flux  remains smaller than 10 percent for the Chandrasekhar-mass
and merger models. The moderate size of the error is due to the
rather small changes in the absolute fluxes compared to the
time-averaged spectra (Figs. 1- 5).
 For the helium-detonation models, the situation is different.  For
this class of models, the large early fluxes and rapid change
with time (e.g. by a factor of 2 between day 14 and 23, Fig. 4)
demand integration times less than $\approx 5 $ days to
achieve the same error level of less than 10 percent.
This should not pose a serious problem because the high
$\gamma$-ray fluxes expected from the helium-detonation models
would allow shorter integration times.
Note that the error in the time of the explosion
can be further reduced if information about the
time evolution of the absolute flux can be determined directly by
$\gamma$-ray observations or, alternatively, by using restrictions
on the model derived by the many constraints outlined in
previous sections.

 \begfig -.2cm
 \psfig{figure=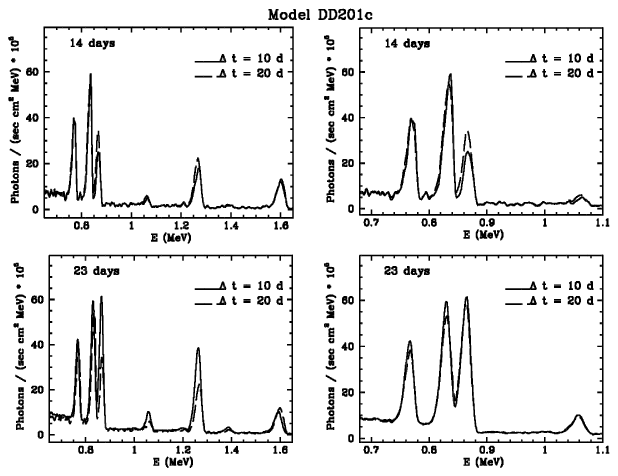,width=13.6cm,rwidth=14.5cm,clip=,angle=270}
\figure{14}
{The predicted spectrum as a function of integration times around day 14 and 23
for the delayed detonation
model DD201c.}
\endfig

\section{Supernova Detection Rate}

The largest obstacle to $\gamma$-ray astronomy of supernovae
is achieving sufficient sensitivity.  This directly affects
the observability of SNe~Ia with the current generation
of $\gamma$-observatories, i.e. CGRO, the upcoming INTEGRAL mission,
and proposed third generation of instruments including Compton and
Laue telescopes.
The sensitivity of the instruments generally increases for intrinsically
narrower lines.  Since the supernova features tend to be
narrower at early epochs (Fig. 7), early time observations are
highly favored for the detection of $\gamma$-ray lines.

The detection limits, $F_{min}$, depend on the mission and instrument.
For the instruments on CGRO, i.e. COMPTEL and  OSSE, $F_{min} =  10^{-4.5}$
photons~sec$^{-1}$ cm$^{-2}$ at about 1 MeV (Johnson et al. 1993).
This limit will unfortunately degrade to about $10^{-4.1}$ photons~sec$^{-1}$
cm$^{-2}$
after the orbit boost that is planned as this paper is written.
According to its design parameters, the goal of INTEGRAL was
to reach $(F_{min}) = 10^{ -5.5}$ photons~ sec$^{-1} cm^{-2}$
(Von Ballmoos et al. 1995).  Recent tests have, however, suggested
that this limit must be revised upward.  Assuming an integration time
of $10^6$ seconds and a 3 $\sigma $ detection, the line sensitivities
are  $10^{ -4.7}$ and $10^{-4.3}$ photons~ sec$^{-1}$ cm$^{-2}$ for
a line width $\delta E/E$ of 1 and 10 \% , respectively.
For the classical $\gamma$-ray lines of \co at 0.84 and 1.24 MeV,
the sensitivity will only be comparable to COMPTEL
(Matteson, 1997). Sensitivities larger by a factor of 2 to 3 can be achieved
at energies of about 100 to about 700 keV.
The continuum sensitivity ranges from about
$10^{-6}$ photons~ sec$^{-1}$ cm$^{-2}$ MeV$^{-1} $ at 100 keV
to $10^{-7.3}$ photons~ sec$^{-1}$ cm$^{-2}$ MeV$^{-1}$ at about 1MeV.

The third generation of proposed  $\gamma$-ray instruments are likely
to use either the Compton effect or Laue refraction.
The first type of instrument reproduces the well-proven technology of COMPTEL.
 The second type uses the new technology
of focussing lenses.
The advantage of the first type is a wide energy range that makes
it the ideal multi-purpose instrument.  The disadvantage is its size which
boosts problems both with the background reduction and with the price tag.
Laue telescopes can, in principle, overcome both of these
disadvantages because of their compactness.  Such instruments are, however,
only able to cover a very narrow energy range comparable to the
intrinsic line widths predicted for supernovae.   Thus they can detect
only one line feature at a time.  Currently, the Laue instruments
must be regarded as an option to study individual selected lines.
For both types of instruments, the realistic design goals are a
broad line sensitivity of the order of $10^{-6}$
photons~sec$^{-1}$ cm$^{-2}$ (Kurfess 1997), substantially more sensitive
than CGRO or INTEGRAL.

The observability of supernovae depends on both the rate of
of occurence of events and their distance.  In recent years,
the number of supernovae discovered nearby has been rather constant,
suggesting that essentially all nearby supernovae that could
be targets for $\gamma$-ray spectroscopy are being discovered.
Table 2 gives the number of SNe~Ia discovered in 1991 and 1992
based on the Asiago catalog.  Only a small fraction of these events
have been analyzed in detail; however, most SNe~Ia are of similar brightness,
comparable to normally bright SNe~Ia.
{}From recent studies, the mean absolute magnitude at maximum in V
is $\approx -19.3^m$ with a spread of about $0.4^m$
(e.g. Hamuy et al. 1996, HK96).
Statistically, the absolute brightness can be translated into an
estimate of distances to individual observed SNe~Ia as presented
in Table 2.

\begtabsmall
\table{2}{Observed SNe~Ia from the Asiago catalog and the
expected $\gamma $-ray flux in the 0.84 MeV line of \co
based on the delayed detonation model DD201c. To get estimates for other models
the results can be scaled according to Fig. 8.
Maximum flux in the model is at 70$^d$.
}
\halign{#\hfil&&\quad#\hfil\cr}
\hline
\+ Max. brightness~~ & No. of disc.~~~~ &expect no.~~~~~& ~~~~appr.
distance~~~~~& 0.84MeV [ph/sec/cm$^2$]\cr
\+ in V             &        1991-92     &  &  [Mpc]              &
$t=70d$~~/~~$t=180d$\cr
\hline
\hline
 \+ 10.0 &    0 &    0.02 &    7.24 &    8.6E-05 & ~~ 4.8E-05 \cr
 \+ 10.5 &    0 &    0.02 &    9.12 &    5.4E-05 & ~~ 3.0E-05 \cr
 \+ 11.0 &    0 &    0.07 &   11.48 &    3.4E-05 & ~~ 1.9E-05 \cr
 \+ 11.5 &    0 &    0.12 &   14.45 &    2.2E-05 & ~~ 1.2E-05 \cr
 \+ 12.0 &    1 &    0.26 &   18.20 &    1.4E-05 & ~~ 7.5E-06 \cr
 \+ 12.5 &    0 &    0.50 &   22.91 &    8.6E-06 & ~~ 4.8E-06 \cr
 \+ 13.0 &    0 &    1.01 &   28.84 &    5.4E-06 & ~~ 3.0E-06 \cr
 \+ 13.5 &    3 &    2.00 &   36.31 &    3.4E-06 & ~~ 1.9E-06 \cr
 \+ 14.0 &    2 &    4.00 &   45.71 &    2.2E-06 & ~~ 1.2E-06 \cr
 \+ 14.5 &    1 &    7.97 &   57.54 &    1.4E-06 & ~~ 7.5E-07 \cr
 \+ 15.0 &    3 &   15.92 &   72.44 &    8.6E-07 & ~~ 4.8E-07 \cr
 \+ 15.5 &    1 &   31.74 &   91.20 &    5.4E-07 & ~~ 3.0E-07 \cr
 \+ 16.0 &    5 &   63.35 &  114.82 &    3.4E-07 & ~~ 1.9E-07 \cr
 \+ 16.5 &    6 &  126.37 &  144.54 &    2.2E-07 & ~~ 1.2E-07 \cr
 \+ 17.0 &    4 &  252.17 &  181.97 &    1.4E-07 & ~~ 7.5E-08 \cr
 \+ 17.5 &    6 &  503.12 &  229.09 &    8.6E-08 & ~~ 4.8E-08 \cr
 \+ 18.0 &   24 & 1003.89 &  288.40 &    5.4E-08 & ~~ 3.0E-08 \cr

\hline
\endtab

{}From the general properties of observed supernovae and the
light curves and spectra of individual events,
the delayed detonation and pulsating delayed detonation models
seem to be the most promising ones to account for the majority
of SNe~Ia events (e.g. H\"oflich et al. 1997  and references therein).
Therefore, we have chosen the delayed detonation model
DD201c to provide a guide line for the observability of SNe~Ia in
$\gamma$-ray lines.  For all models of normally bright events,
the maximum $\gamma$-ray flux varies only within $\approx$
50 percent (see Fig. 8).  Since the flux scales inversely with the
distance that effect will dominate, and our particular choice
of model does not substantially effect this discussion.  For SNe~Ia at the
luminous end
of the observed luminosity distribution, distances up to
which a detection may be possible should be scaled up by,
at most, 50 percent.

For CGRO after the orbit boost with a detectability limit of $F_{min} \approx
10^{-4.1}$
photons sec$^{-1}$ cm${-2}$,
Table 2 shows that normally bright SNe~Ia will be detectable
up to a distance of 7 Mpc with an upper limit of about 10 Mpc.
Statistically, there is an SNe~Ia at a distance between 4 and 9 Mpc
about once every 8 years (e.g. 4.6 Mpc for SN1972e,
4.5 Mpc for SN1986g, 8.4 Mpc for SN1989b) somewhat higher than the simple
estimate of Table 2.
 Such an event would clearly allow
for a positive measurement of the $\gamma$-ray fluxes or would rule out
all the present models.  Due to the long expected operational time of
CGRO ($\geq $ 10 years), there is a good chance that a clear
detection of \co lines could be obtained which would allow
different models within each explosion scenario to be distinguished.
The low energy resolution of CGRO would not permit the use of
information on the velocity distribution. If, however, the relative
fluxes of, e.g., the  0.84 MeV and 1.24 MeV lines could be measured
to within an accuracy of 10 percent and observations are available
up to 50 days after the explosion, the line ratios would provide a
distinction between the explosion of a massive white dwarf and that
of a sub-Chandrasekhar, helium-detonation.

A much cleaner distinction between Chandrasekhar and sub-Chandrasekhar models
 could be achieved if observations were
available at early epochs, about 15 to 20 days after the
explosion corresponding to about maximum visual light.
Several SNe~Ia have been discovered up to about 2 weeks before maximum
(see e.g. Hamuy et al. 1995 for a list), so $\gamma$-ray observations
at these times may not be unrealistic.
 Figure 10 shows that the peak \ni flux of the sub-Chandrasekhar models exceed
that of the Chandrasekhar mass models by a factor of 5 for normally bright
models
about 10 days after the explosion and by factors of several at later epochs. At
10 days
after the explosion, the normally bright helium detonation models predict a
flux  of
 $\approx 2x10^{-4}$ photons sec$^{-1}$ cm$^{-2}$ at 10 Mpc, and at maximum
light,
$\approx 15 $ days after the explosion, a flux of  $\approx 10^{-4}$ photons
sec$^{-1}$ cm$^{-2}$,
factors of about 5 and 2, respectively, brighter than predicted at 70 days.
Given the high maximum
\ni line fluxes of helium-detonation models, both normally bright
and subluminous Type Ia should be detectable by CGRO up to distances of
15 and 10 Mpc, respectively, if the observations start about
 1 week before maximum light in the optical wavelength range or these scenarios
can be ruled out.
SNe~Ia within 15 Mpc occur about once every 5-10 years.
In particular, the helium-detonation scenario  predicts the
detectability by CGRO of all normally bright SNe~Ia before optical
maximum in the Virgo cluster at about 18-20 Mpc where an SNe~Ia should have an
apparent magnitude of V=12-12.5.
Based on Table 2, the opportunity for such an observation
might be available every 3 to 4 years as supernova searches of all kinds
become more successful at discovering SNe~Ia well before maximum
light.  In particular, SN 1991T would have provided a rigorous
test of the helium-detonation models had the CGRO observations
been made at a sufficiently early epoch.
 The key point is thus if an ppropriately early CGRO observation significantly
fails
to detect a flux of $\approx 5x10^{-5}$ photons sec$^{-1}$ cm$^{-2}$ from an
SNe~Ia
of V=12.5, then helium detonation models can be ruled out. A detection or even
a firm
upper limit for any SNe~Ia in the Virgo Cluster will provide a much needed
distinction
between Chandrasekhar and sub-Chandrasekhar models.

With INTEGRAL, the next generation $\gamma$-ray observatory,
all the types of observations discussed here are feasible for the
classical strong \co lines which have been discussed
in the context of CGRO.  In addition, the high velocity resolution
($\approx 600 $ km sec$^{-1}$, Matteson 1997) will allow detailed
studies of line profiles (see above). An especially exciting possibility
arises from the high sensitivity at low energies and the low
energy threshold of INTEGRAL which will allow detections of
 both the 0.511 MeV line of positronium
and the \ni line at 0.48 MeV.  Although the corresponding lines are
weaker by a factor of 4 compared to the strongest lines of \ni and \co,
the increased low energy sensitivity of INTEGRAL should allow detection
of the positron decay line up to about 10 to 13 Mpc for normally
bright SNe~Ia over a period of several months.
Observations during the first month after explosion would allow detailed
studies of \ni lines and the 0.511 MeV line and, consequently,
would provide for the first time a realistic way to determine the
time of the explosion (Fig. 13).
 With INTEGRAL, the very strong \ni lines at about 0.16 and 0.27
and the Compton continuum can be detected (Fig. 1) out to
distances including the near side of the Virgo cluster.
Therefore, a target of opportunity could be expected every 3 to 4 years.

For the third generation instruments, all ~SNe~Ia within a
distance of 70 to 100 Mpc would be within reach.
This translates into about 6 targets of opportunity per year.
Detailed studies of a variety of supernovae including their
statistical properties would be possible.  All the effects mentioned
in the previous sections could be studied in great detail,
opening a new window to do $\gamma$-ray astronomy of supernovae
on a regular basis.

The sensitivity of the proposed third-generation experiments brings
an entirely new aspect to $\gamma$-ray astronomy of supernovae.
Up to this point, the current discussion has been based on the premise
that all supernovae to be studied will be discovered by standard
ground-based optical search techniques.  The record of distant
supernovae discovered in this way must be very incomplete.
As can be seen from Table 2, the number of discoveries does
not scale with the volume, even for distances not effected by clustering.
This strongly suggests the presence of selection effects
which must depend on the Type of the host galaxy, the extinction,
and the distance from the center of the galaxy.
These questions are highly relevant to clarify the progenitor
population and its influence on the rate of discovery and nature
of the supernovae. These issues have important consequences for
the use of SNe~Ia to obtain extragalactic distances at high red shift
(H\"oflich et al.  1997).


\section{Positron Trapping}

One of the more interesting and still unsolved problems of
SNe~Ia is the physics that determines the slope of the
late-time light curve.  As emphasized by Colgate (Colgate,
Petschek \& Kriese 1980, Colgate 1983, Colgate et al. 1997), the
exponential decay continues for many e-folds, in excess
of 600 days (Kirshner and Oke 1972, Kirshner et al 1993).
There is no observational evidence for the decline of the optical light
curve due to an ``infrared catastrophe,"  (Weaver, Axelrod  \& Woosley 1980,
Fransson et al 1996) that is predicted to occur as standard
thermonuclear models expand and cool.  With assumed complete
trapping of positrons due to tangled magnetic fields and
the resulting small Larmor radii, the line emission
shifts to the IR fine structure lines of iron.  Colgate has
argued that this problem requires a complete shift of
paradigm and in favor of a model based on core-collapse
that ejects only a small amount of highly nickel-enriched matter
($\sim 0.4 M_\odot$)
Colgate proposes that with the small mass and associated radial combing
of the magnetic field, the positrons would leak out with a
deposition function that is essentially self-similar to that
for $\gamma$-rays.  He argues that the resulting deposited
power in positrons could be restricted to optical emission
through fluorescence effects and could then give the observed
long time exponential decay.   This model is very unlikely to
apply to SNe~Ia since it would give spectra anomolously rich
in \ni and poor in intermediate mass elements, continuua that
are too blue (for the same reasons as the helium-detonation
models), excessively high velocities, light curves that
peak and decline too rapidly, and implied distances
that are incommensurate with those determined by
$\delta$ Cepheids.  Nevertheless, Colgate is right that
the problem of the late-time decay must be solved
in the context of the thermonuclear models before they
can be regarded as complete.  Gamma-ray observations
should help to clarify this important issue of physics.

To constrain positron trapping, late-time observations
are necessary.  Even with no magnetic field, positrons
are not predicted to begin leaking, thereby reducing
the positron energy deposition function, until about
300 days.  The flux is correspondingly small at this
time.  INTEGRAL would require a supernova within
only a few megaparsecs to make the corresponding detection.
The next generation of $\gamma$-ray telescopes could,
however, make a substantial contribution to this issue.
A flux limit of $10^{-6}$ photons~ sec$^{-1}$ cm$^{-2} $
would allow a measurement of the difference in the positron
flux between fully trapped and freely leaking models
for supernovae out to the Virgo Cluster.

The issue of the positron leakage could be approached in
several ways, some of which might be distance dependent
and some of which would require an independent measure
of the distance.  One method would be to look carefully
at the 0.511 Mev line flux, in an absolute sense, or
in comparison to the \co lines at the same epoch.
Another technique, which would not require narrow line
resolution would be to look at the absolute or relative photon
flux of the positronium continuum.
 Because the photon flux is constant
it may even be detectable as a constant contribution to the the X-ray flux by
   X-ray instruments due to their superior  sensitivity.

\section {Discussion and Conclusions}

We have presented $\gamma$-ray spectra and light curves for a variety of
models of SNe~Ia including delayed detonation, pulsating
delayed detonations, merger scenarios and helium detonations.
Gamma rays trace the radioactive isotopes whereas optical
photons trace elements and, therefore, $\gamma $ observations
must be seen as an important complement to those at other wavelengths.

A detailed discussion of the properties of the emitted spectra were given.
These properties provide sensitive tools to study details of
the explosion models and information which is relevant for the data analysis
of the $\gamma$-ray observations themselves.
Since the different model scenarios show very different signatures
in their time evolution, spectral features and line ratios,
the observation of $\gamma$-rays have the capacity to remove any
ambiguity about the basic nature of Type Ia supernovae, whether
they involve thermonuclear explosion or core collapse and in
the expected former case, whether they involve Chandrasekhar or
sub-Chandrasekhar mass white dwarfs.
For nearby  supernovae, $\gamma$-ray spectroscopy has the potential
to separate effects due to the initial metallicity of the exploding white
dwarf, mixing, and nuclear burning.

A comparison of $\gamma$-ray fluxes and bolometric (UV/optical/IR)
light curves can provide a measure of the escape probability of positrons
and, in principle, of the magnetic field strength and distribution.
Gamma-rays provide an unbiased, direct tool to test
the calibration of SNe~Ia and to overcome systematic effects due
to reddening  and errors in the calibration of secondary
distance indicators. Such tests are critical for the future use of
SNe~Ia as distance indicators to determine the deceleration parameter
and other cosmologically  relevant parameters.

SNe~Ia are not a strictly homogeneous group of objects as
previously believed. Therefore, any study of them must include a
large number of events. For $\gamma$-ray astronomy, this implies
that a broad-line sensitivity of about $10^{-6}$ photons s$^{-1}$ cm$^{2}$
is needed to bring within reach all SNe~Ia up to a distance of 70 to 100 Mpc.
Based on current discovery rates in the optical, this translates
into about 6-8 SN yr$^{-1}$ each of which would be detectable for a period of
about 6 months.

The increased sensitivity of future generations of
instruments will give for the first time the absolute
and the relative rate of SNe~Ia as a function of the galaxy type
unbiased by extinction and the type of the host galaxy which, as can be seen
from Table 2, hampers optical searches.
A monitoring of distant clusters
on a regular basis is  feasible, because an entire cluster can be
observed simultaneously given  the wide field of view of gamma ray detectors.
  This should give valuable information on the progenitor evolution,
still one of the major mysteries associated with SNe~Ia.

\noindent
{\bf Acknowledgements:} PAH wants to thank J.D. Kurfess for the organization of
the
excellent workshop on Gamma Ray Astrophysics in Washington
and all participants for the interesting discussions.
 This research was supported in part by NSF Grant AST 9528110, NASA Grant
NAG 5-2888 and by a grant from tahe Texas Advanced Research Program.
\bigskip
\heading{References}

\journal{ Ambwani K., Sutherland P.G.}{ 1988}{ ApJ}{ 325}{ 820}

\inbook
 {Arnett D.}
{1997}{Thermonuclear Supernovae}{P. Ruiz-Lapuente, R. Canal \& J. Isern}
{Kluwer Academic Publishers}{Dordrecht/Boston/London}{405}
\journal{ Benz W., Bowers R.L., Cameron A.G.W., Press W.H.}{ 1989}{ ApJ}{ 325}{
820}

\journal{ Burrows A., The, L.-H.}{ 1990}{ ApJ}{ 360}{ 626}

\inbook{ Chan K.W., Lingenfelter R.E.}{ 1988}
{ Nuclear Spectroscopy
ot Astrophysical Sources}
{eds. N. Gehrels, G. Share}{AIP}{New York}{  110}

\journal{ Chan K.W., Lingenfelter R.E.}{ 1990}{ Proc. 21st Internat.
Cosmic Ray Conference, Adelaide}{ 1}{ 101}

\journal{Chan K.W., Lingenfelter R.E.}{1991}{ApJ}{ 368}{ 515}
\journal{ Clayton D.D., Colgate S.A., Fishman G.J.}{ 1969}{ ApJ}{ 155}{ 75}

\journal {Clayton D.D.}{1974}{ ApJ}{ 188}{ 155}

\journal {Colgate S.A., Peschek A.G., Kriese J.T.}{ 1980}{ApJ}{320}{304}

\inbook{ Colgate S.A.}{ 1983}
{ Positron-Electron Pairs in Astrophysics}
{Greenbelt, MD}{American Institute of Physics}{New York}{94}

\inbook
 {Colgate S.A., Fryer C.L., Hand, K.P.}
{1997}{Thermonuclear Supernovae}{P. Ruiz-Lapuente, R. Canal \& J. Isern}
{Kluwer Academic Publishers}{Dordrecht/Boston/London}{273}

\journal{Collela  P., Woodward  P.R.}{1984}{J.~Comp.~Phys.}{54}{174}

\journal{Freedman W.L. et al.}{1994}{ApJ}{457}{628}

\journal {H\"oflich  P., Khokhlov A., M\"uller E.} {1992} {A\&A} {259} {549}

\journal {H\"oflich  P., Khokhlov A., M\"uller E.} {1993a} {A\&A} {268}{470}

\journal {H\"oflich  P., Khokhlov A., M\"uller E.} {1993b} {A\&A Suppl.} {97}
{221}

\journal {H\"oflich  P., Khokhlov A., M\"uller E.} {1994} {ApJ Suppl.} {92}
{501}

\journal {H\"oflich  P.} {1995} {ApJ} {443} { 89}

\journal {H\"oflich  P., Khokhlov A., Wheeler J.C } {1995} {ApJ} {444} { 831}

\journal {H\"oflich  P., Khokhlov A.} {1996} {ApJ} {457} {500}

\inbook
 {H\"oflich P. , Khokhlov A., Nomoto K., Thielemann F.K., Wheeler J.C.}
{1997}{Thermonuclear Supernovae}{P. Ruiz-Lapuente, R. Canal \& J. Isern}
{Kluwer Academic Publishers}{Dordrecht/Boston/London}{659}

\journal{ Hubbell J.H.}{ 1969}{ NSRDS}{NBS}{ 29}

\journal{Johnson W.N. et al}{ 1993}{ ApJS}{86}{693}

\journal {Khokhlov  A.}{1991}{A\&A} {245} {114}

\journal {Khokhlov  A., M\"uller E., H\"oflich P.}{1992}{A\&A} {253} {L9}

\journal {Khokhlov  A., M\"uller  E.,   H\"oflich P.} {1993} {A\&A} {270} {223}

\journal{Khokhlov A.}{1996}{ApJ}{447}{L73}

\infuture{Khokhlov A., Oran E.S., Wheeler J.C}{1997a}{ApJ}{in press}

\infuture{Khokhlov A., Oran E.S., Wheeler J.C}{1997b}{Combustion and Flame}{in
press}

\journal
 {Kirshner R.P., Oke J. B.}
{1975}{ApJ}{200}{574}

\journal
 {Kirshner R.P. et al.}
{1993}{ApJ}{415}{589}

\inbook
 {Kumagai K, Nomoto K}
{1997}{Thermonuclear Supernovae}{P. Ruiz-Lapuente, R. Canal \& J. Isern}
{Kluwer Academic Publishers}{Dordrecht/Boston/London}{515}

\infuture
{Kurfess J.D.} {1997}
{Workshop on Low/Medium Energy Gamma Ray Astrophysics Missions}
{Landsdown, Washington, November 1996}

\journal
 {Leising M. D., Johnson W. N.,  Kurfess J. D., Clayton D. D.,  Grabelsky D.
A., Jung G. V.,
Kinzer R. L., Purcell W. R.,  Strickman M. S., The L. S.,  Ulmer M. P.}
{1995}{ApJ}{450}{L805}

\journal {Lichti  G. G.,Bennett  K., Herder  J. W. D.,Diehl  R., Morris
d.,Ryan  J.,
Schoenfelder  V.,Strong  A. W., Winkler  C.S, Winkler  C.}{1994} {A\&A}
{292}{569}

\journal {Livne E., Glasner A.S.}{1990}{ApJ}{361}{L244}

\journal {Livne E., Arnett, D.}{1995}{ApJ}{452}{L62}

\infuture
{Matteson M.} {1997}
{Workshop on Low/Medium Energy Gamma Ray Astrophysics Missions}
{Landsdown, Washington, November 1996}

\journal {Meikle, W.P.S. et al.}{1996}{MNRAS}{281}{263}

\journal {Niemeyer J.C. Hillebrandt W.} {1995} {ApJ} {452}{769}

\journal {Niemeyer J.C. Woosley S.E.} {1997} {ApJ} {475}{740}

\journal {Nomoto  K., Sugimoto  D.}{1977}{PASJ}{29}{ 765}

\inbook {Nomoto K.} {1980} {IAU-Sym. 93}{D. Sugimoto, D.Q. Lamb \& D.
Schramm}{Dordrecht}{Reidel}{295}

\journal { Nomoto  K.}{1982}{ApJ}{253}{ 798}

\journal{Nomoto  K., Thielemann  F.-K., Yokoi  K.}{1984}{ApJ}{286}{ 644}

\infuture {Nugent P., Baron E., Branch D., Hauschildt P. H.}
{1997}{ApJ}{in preparation}

\journal {Perlmutter S. et al.}
{1995}{ApJ}{440L}{41}

\journal {Riess A.G., Press W.H., Kirshner R.P.}
{1996}{ApJ}{473}{588}

\journal {Ruiz-Lapuente  P., Jeffery D., Challis P.M., Filippenko  A.V.,
Kirshner R.P., Ho L.H., Schmidt B.P., Sanchez F., Canal R.
}{1993}{Nature}{365}{728}

\journal {Saha  A., Sandage  A., Labardt  L., Tammann  G. A.,  Macchetto  F.
D., Panagia  N.}
{1996}{ApJS}{107}{693}

\journal {Sandage A., Tammann G.A.}
{1996}{ApJ}{464L}{51}

\infuture { Schmidt, B. et al.}{1997}{BAAS}{in press}

\inbook{Thielemann  F.-K., Nomoto  K., Hashimoto  M.}{1994}
{Supernovae}{ Les Houches}{S. Bludman  R., Mochkovitch  J.,
Zinn-Justin}{Elsevier, Amsterdam}{629}

\journal {Veigele W.J.}{1973}{Atomic data tables}{5}{51}

\inbook{Von Ballmoos  P., Dean T., Winkler  C.}{1995}{$17^{th}$ Workshop on
Relativistic Astrophysics}
{H. B\"ohringer, G.E. Morfill \& J.E. Trumper}{Ann. of the New York Academy of
Science}{Vol. 759}
{401}

\inbook{Weaver T.A., Axelrod T.S., Woosley S.E.}{1980}{ Type I Supernovae}
{J. C. Wheeler}{Austin}{Texas}{113}

\inbook{Wheeler J. C}{1996}{Evolutionary Process in Binary Stars}
{R. A. M. Wijers, M. B. Davies, \& C. A. Tout}{Dordrecht}{Kluwer}{307}

\inbook {Woosley S. E., Weaver T.A., Taam R.E. } {1980}
{in: Type I Supernovae}{J. C. Wheeler}{Austin}{U.Texas}{96}

\journal {Woosley  S. E. \& Weaver, T. A.} {1994}{Ap. J.} {423}{371}

\journal {Yamaoka H., Nomoto K., Shigeyama T., Thielemann
F.}{1992}{ApJ}{393}{55}

\end